\documentclass[aps,prl,amsmath,amsfonts,floatfix,superscriptaddress,footinbib,twocolumn]{revtex4-1}
\usepackage[colorlinks=true]{hyperref}
\hypersetup{
  colorlinks=true,
  linkcolor=blue,
  filecolor=magenta,
  citecolor=blue,
  urlcolor=blue,
}

\usepackage{graphicx,color}
\usepackage{bbm}
\usepackage{mathrsfs} \usepackage{amssymb} 
\usepackage{commath}\usepackage{tikz}
\usetikzlibrary{shapes,arrows,positioning,calc}\usepackage{stmaryrd}\usepackage{bm}

\def\vr{\Vec{r}}
\def\vR{\Vec{R}}
\def\vu{\Vec{u}}
\def\veta{\bm \eta}
\def\vS{\Vec{S}}

\newcommand{\ov}[1]{\overline{#1}}
\newcommand{\ovS}{\overline{\Vec{S}}}
\newcommand{\ovr}{\overline{\Vec{r}}}
\newcommand{\ovu}{\overline{\Vec{u}}}

\newcommand{\eq}[1]{Eq.~(\ref{#1})}
\newcommand{\beq}{\begin{equation}} \newcommand{\eeq}{\end{equation}}
\newcommand{\beqn}{\begin{eqnarray}} \newcommand{\eeqn}{\end{eqnarray}}

\newcommand{\cD}{{\cal D}}

\renewcommand{\Vec}[1]{{\bf #1}}

\newcommand{\intS}{\int_{S}}
\newcommand{\intr}{\int d^{d}r}

\begin{document}

\title{Translational and orientational glass transitions in the large-dimensional limit : a generalized replicated liquid theory and
  an application to patchy colloids}

\author{Hajime Yoshino}
\affiliation{Cybermedia Center, Osaka University, Toyonaka, Osaka 560-0043, Japan}
\affiliation{Graduate School of Science, Osaka University, Toyonaka, Osaka 560-0043, Japan}
 \email[⟨Email:⟩]{yoshino@cmc.osaka-u.ac.jp}

\begin{abstract}
  We developed a generalized replicated liquid theory for glassy phases of non-spherical uniaxial molecules and colloids, which becomes exact in the large dimensional limit $d\to\infty$. We then applied the scheme to patchy colloids with sticky patches at their heads and tails. The system exhibits rich phase behaviors involving the translational and orientational degrees of freedom.  We found a novel glass-glass transition between glasses with large/small orientational fluctuations.
  \end{abstract}

\maketitle

\paragraph*{Introduction --}

Understanding properties of glasses from 1st principles is a major challenge in physics. From theoretical point of view, a sensible strategy would be to start from the simplest glass forming system and progressively approach diverse real systems step by step. The simplest glass forming system is the assembly of hardspheres in the limit $d \to \infty$ which has been exactly analyzed recently
from the emergence of glasses up to jamming using the replicated liquid theory \cite{parisi2010mean,kurchan2012exact,kurchan2013exact,charbonneau2014exact,charbonneau2014fractal}.

Since molecules and colloids are generically non-spherical, a natural step forward to go beyond the simplest spheres is to consider particulate systems with orientational degrees of freedom. An important question is whether  the orientational degrees of freedom is essentially subjected to the translational one or they can play distinct roles in glass transitions \cite{suga1974thermodynamic,letz2000ideal,chong2002idealized,chong2005evidence,de2007dynamics,zheng2011glass,mishra2013two} and in glasses. To tackle this problem from 1st principles we develop a replicated liquid theory for a class of generic uniaxial particles
taking into account both the translational  \cite{mezard1999first,parisi2010mean,kurchan2012exact,kurchan2013exact,charbonneau2014exact} and the orientational degrees of freedom  \cite{yoshino2018} in the limit $d \to \infty$ to establish an exact mean field statistical mechanics framework.

\begin{figure}[h]
   \includegraphics[width=0.48\textwidth]{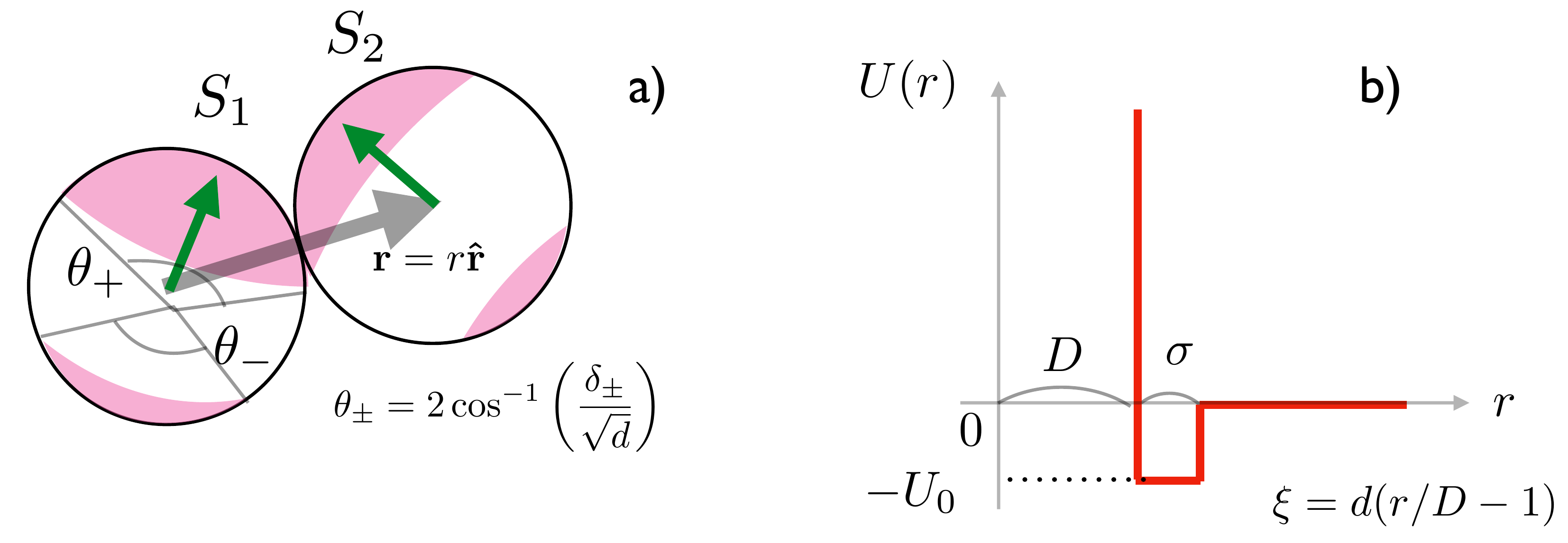}
 \caption{ (Left) Patchy colloids with two sticky patches at their heads(+) and tails(-).
   The size of the patches can be parameterized
   by the visual angle $\theta_{+}$ and $\theta_{-}$. The green arrows indicate the directors.
   (Right) Square-well potential.
 }
  \label{fig_patchy_colloid}
\end{figure}

One of the simplest particulate systems with translational/rotational degrees of freedom is an assembly of uniaxial patchy colloids which are geometrically spherical but have orientational degrees of freedom (see Fig.~\ref{fig_patchy_colloid}). Patchy colloids have attracted significant interests because of the possibilities to design valence limitted crystalline  \cite{chen2011directed}
and amorphous structures \cite{sciortino2008gel,bianchi2006phase,ruzicka2011observation}. In the limit that the patches fully cover the surfaces they become reduced to the standard sticky colloids for which interesting phase behaviors such as reentrant glass transition and repulsive/attractive glass transitions are known \cite{bergenholtz1999nonergodicity,fabbian1999ideal,pham2002multiple,eckert2002re,pham2006yielding,sciortino2002disordered,sellitto2013thermodynamic}.
In this work, we uncover richer phase behaviors in the presence of the rotational degrees of freedom including novel glass-glass transition driven by the orientational degrees of freedom.

\paragraph*{Exact glass free-energy functional for
an assembly of generic uniaxial particles in $d \to \infty$ limit}

Let us consider an assembly of {\it anisotropic} particles with axial symmetry
in the $d$-dimensional space.
The configuration of the particles $i=1,2,\ldots,N$
is specified by translational coordinates
$\vr_{i}=(r_{i}^{1},r_{i}^{2},\ldots,r_{i}^{d})$ 
and spins or directors  $\vS_{i}=(S^{1}_{i},S^{2}_{i},\ldots,S^{d}_{i})$
which are normalized such that $\sum_{\mu=1}^{d}(S^{\mu})^{2}=d$.
The particles interact with each other through a two-body potential,
\beq
H=\sum_{i < j} V(\vr_{ij},\vS_{i},\vS_{j}).
\eeq
In the following the specific shape of the potential is not important but
we assume orientational and translational
invariance such that it can be parameterized as,
\beq
V(\vr_{12},\vS_{1},\vS_{2})=
V(r_{12},\hat{\bf r}_{12}\cdot \vS_{1},\hat{\bf r}_{12}\cdot \vS_{2},\vS_{1}\cdot\vS_{2})
\eeq
where $\vr_{12}=\vr_{1}-\vr_{2}$, $\hat{r}=\vr/r$ and $r=|\vr|$.

In order to study glass transitions, we develop a replicated liquid theory which becomes exact
in the large dimensional limit $d\to \infty$ limit.
This amount to consider a liquid of 'molecules' $i=1,2,\ldots,N$
made of $m$ replicas.  \cite{parisi2010mean,kurchan2012exact}
The $i$-th molecule consists of replicas $a=1,2,\ldots,m$ 
whose translational/orientational coordinates can be written as
$\vr^{a}_{i}=\vR_{i}+(D/\sqrt{d})\veta_{i}^{a}$ and $\vS^{a}_{i}$.
Here $D$ is the size of the particle and $\vR_{i}$
is the center of mass coordinate of the 'molecule'.
The glass order parameter for
the translational degree of freedoms \cite{kurchan2013exact} is
$\alpha_{ab}=\langle \veta^{a}_{i} \cdot \veta^{b}_{i} \rangle$
which is related to the 'cage size' $\Delta_{ab}=\alpha_{aa}+\alpha_{bb}-2\alpha_{ab}$. The latter is infinite in liquids
but becomes finite $\Delta_{ab} < \infty$ ($a\neq b$)
in glasses of the translational degrees of freedom.
We introduce the orientational glass order parameter  as
$Q_{ab}=(1/Nd)\sum_{i=1}^{N}\langle \vS^{a}_{i} \cdot \vS^{b}_{i} \rangle$,
which is zero in liquids
but becomes non-zero $Q_{ab} > 0$  ($a\neq b$)
in glasses of the rotational degrees of freedom.
We also introduce $\beta_{ab}=1/(N\sqrt{d})\sum_{i=1}^{N}\langle \veta^{a}_{i} \cdot \vS^{b}_{i} \rangle$ which represents the cross-correlations.

The free-energy functional of the replicated
system in the $d \to \infty$ limit is obtained as (See
\footnote{Supplemental Material} for the details),
\begin{eqnarray}
  -\beta m \phi_{m}[\hat{\cal Q}] =
      c_{\rm nt}
 +   \frac{d}{2}\ln {\rm det} \hat{\cal Q}^{2m,2m}
 -\frac{d}{2}\hat\varphi {\cal F}_{\rm int}
 [\hat{\cal Q}] \qquad
 \label{eq-free-ene}
\end{eqnarray}
with $c_{\rm nt}=
1-\ln \rho  + d \ln m
    +\frac{(m-1)d}{2}\ln \left(\frac{2\pi eD^{2}}{d^{2}}\right)
    +\frac{d}{2}m\ln \left( \frac{2\pi e}{d}\right)$.
Here $\beta=1/k_{\rm B}T$ is the reduced temperature
(not to be confused with the order parameter $\hat{\beta}$),
$\rho$ is the number density, $\hat\varphi=2^{d}\varphi/d$
is the reduced volume fraction with $\varphi$ being the volume fraction.
For convenience we have introduced a matrix
$\hat{{\cal Q}}$ which is a super-matrix of size $2m \times 2m$ consisting of
$m\times m$ diagonal sub-matrices $\hat{Q}$ and $\hat{\alpha}$,
off-diagonal ones $\hat{\beta}$ and $\hat{\beta}^{\rm t}$ \cite{Note1}.
We also introduced a matrix $\hat{{\cal Q}}^{2m,2m}$ which 
is defined by subtracting the $2m$-th row and column of $\hat{{\cal Q}}$.
The 1st and 2nd terms on the r.h.s. of  \eq{eq-free-ene}
can be regarded as the entropic part of the free-energy while
the last term can be regarded as the interaction part.
The functional ${\cal F}_{\rm int} [\hat{\cal Q}]$ is given by,
\begin{eqnarray}
&&   -{\cal F}_{\rm int}[\hat{\cal Q}]=
  \int_{-\infty}^{\infty} d\xi e^{\xi}   \left. e^{
    \frac{1}{2}\sum_{a,b=1}^{m} {\cal D}_{ab}
        }  f_{m}(\ov{\xi},\ov{x},\ov{x'},\ov{h})\right|_{\substack{ \{\ov{\xi}=\xi \nonumber \\ \ov{x}=\ov{x'}=\ov{h}=0\} }}
  \nonumber   \\
&&      {\cal D}_{ab}
    \equiv 
    -\Delta_{ab}\partial_{\xi_{a}}\partial_{\xi_{b}}
    +Q_{ab} (
         \partial_{x_{a}} \partial_{x_{b}}
        +\partial_{x'_{a}}\partial_{x'_{b}})
        +(Q_{ab})^{2}\partial_{h_{a}}\partial_{h_{b}}  \nonumber \\
&&     f_{m}(\ov{\xi},\ov{x},\ov{x'},\ov{h}) \equiv
       \prod_{a=1}^{m}  e^{-\beta V(D(1+\frac{\xi_{a}}{d}),x_{a},x'_{a},h_{a})}
         -1
     \label{eq-f-int-anisotropic-particle}
\end{eqnarray}
where we introduced a compact notation
$\overline{x}=\{x_{1},\ldots,x_{m}\}$.
  The function $f_{m}(\ov{\xi},\ov{x},\ov{x'},\ov{h})$
  is the replicated Mayer function.
Note that ${\cal F}_{\rm int}[\hat{\cal Q}]$ depends on $\Delta_{ab}$, $Q_{ab}$
but not on $\beta_{ab}$ respecting the translational/rotational invariance. 

\paragraph*{1 step RSB ansatz --}

The simplest ansatz for the glass order parameters
is the 1 step replica symmetry breaking (1RSB) ansatz for which we have
$Q_{ab}=(1-q)\delta_{ab}+q$,
$\beta_{ab}=(m\delta_{ab}-1)\beta$ $\alpha_{ab}=(m\delta_{ab}-1)\alpha$
and $\Delta_{ab}=\Delta(1-\delta_{ab})$ with $\Delta=2m\alpha$.
The free-energy functional becomes,
\begin{eqnarray}
&&   -\beta m  \phi_{m}(\Delta,q,\beta) = c_{\rm nt}   \\
&&   +   \frac{d}{2}\left[
    (m-1)\ln\left(\frac{\Delta}{2}-\frac{m^{2}\beta^{2}}{1-q}\right)-\ln m  \right]
 + \frac{d}{2}\hat\varphi {\cal F}_{\rm int}(\Delta,q)    \nonumber
   \label{eq-replica-free-energy-1RSB}
\end{eqnarray}
with
\beqn
    {\cal F}_{\rm int}(\Delta,q)&=&
     \int_{-\infty}^{\infty}d\xi e^{\xi}
 \biggl( \int {\cal D}z_{x}{\cal D}z_{x'}{\cal D}z_{h} \nonumber \\
&&  \left.g^{m}(\Xi) \right |_{\Xi=(\xi+\Delta/2,\sqrt{q}z_{x},\sqrt{q}z_{x'},\sqrt{q^{2}}z_{h})}  -1 \biggr) \qquad
 \label{eq-Fint-1RSB-v2}
\eeqn
with
\beqn
&& g(\xi,x,x',h)
= \int {\cal D}z_{\xi} {\cal D}z_{x}{\cal D}z_{x'}{\cal D}z_{h} \nonumber \\
&& e^{-\beta V(\xi-\sqrt{\Delta}z_{\xi},x-\sqrt{1-q}z_{x},x'-\sqrt{1-q}z_{x},h-\sqrt{1-q^{2}}z_{h})} \qquad
\label{eq-g-integral-form}
\eeqn
Around $m=1$ it is useful to expand the free-energy in power series
of $s=1-m$ which yields,
\beq
-\beta m \phi_{m}(\Delta,q,\beta)=
-\beta \phi_{1}-\beta \nu_{\rm FP}(\Delta,q,\beta)s+O(s^{2})
\label{eq-def-FP}
\eeq
where $\nu_{\rm FP}(\Delta,q,\beta)$ is the so called Franz-Parisi potential.

The values of the order parameters $\Delta$, $q$ and $\beta$ should be
obtained by solving the saddle point equations which extremize the free-energy function. Apparently the cross-correlation vanishes $\beta=0$. After fixing the remaining order parameters $q$ and $\Delta$, we are left with the replica free-energy $\phi_{m}=\phi_{m}(\Delta^{*},q^{*})$ which is still a function of the parameter $m$. Then following the standard prescription \cite{Mo95}, various
thermodynamic quantities including the complexity or the
configurational entropy $\Sigma(m)$
can be obtained  \cite{Note1}. By increasing the density $\hat\varphi$
with $m=1$ we would meet a so called
dynamical transition density $\hat\varphi_{\rm d}$
above which non-trivial solutions with $\Delta^{*} < \infty$ and/or $q^{*} >0$
emerge. The thermodynamic glass transition, called as the
Kauzmann transition would take place at a higher density $\hat\varphi_{\rm K}$
where the complexity vanishes $\Sigma(1)=0$.
Let us call the regime $\phi_{\rm d} < \phi < \phi_{\rm K}$
as {\it glassy liquid} regime since $\Sigma(1) >0$ suggests
that the system is still in the liquid state
but exhibits glassy dynamics hopping between different free-energy
basins. In the ideal, thermodynamic glass state
which would exist at higher densities $\varphi_{\rm K} < \varphi$
the parameter $m$ should be set
at $m=m^{*}(\hat\varphi)$ so that the complexity remain zero, i~.e.
$\Sigma(m^{*})=0$ and it would decrease with increasing density.
Eventually at the glass close packing density $\hat\varphi_{\rm GCP}$
where the ideal glass state exhibits jamming, $m^{*}$ vanishes.

\begin{figure*}[t]
    \includegraphics[width=0.9\textwidth]{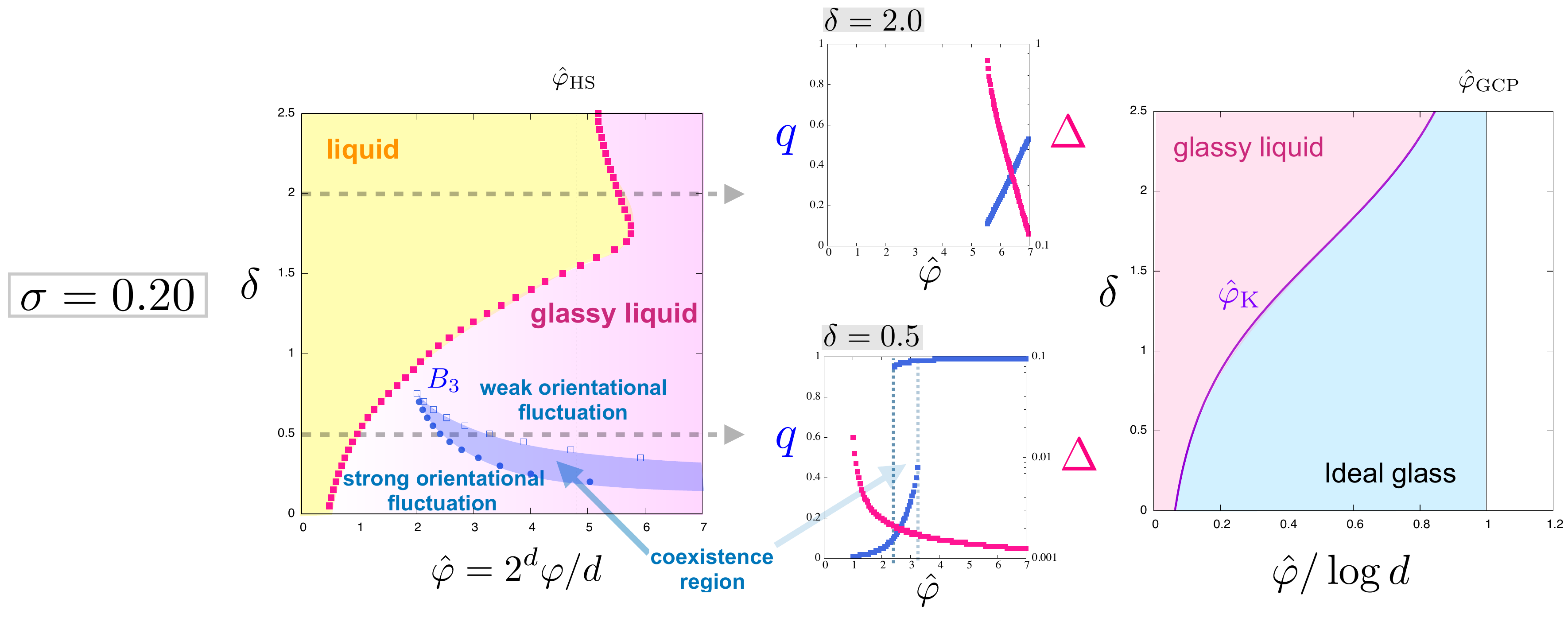}
   \includegraphics[width=0.9\textwidth]{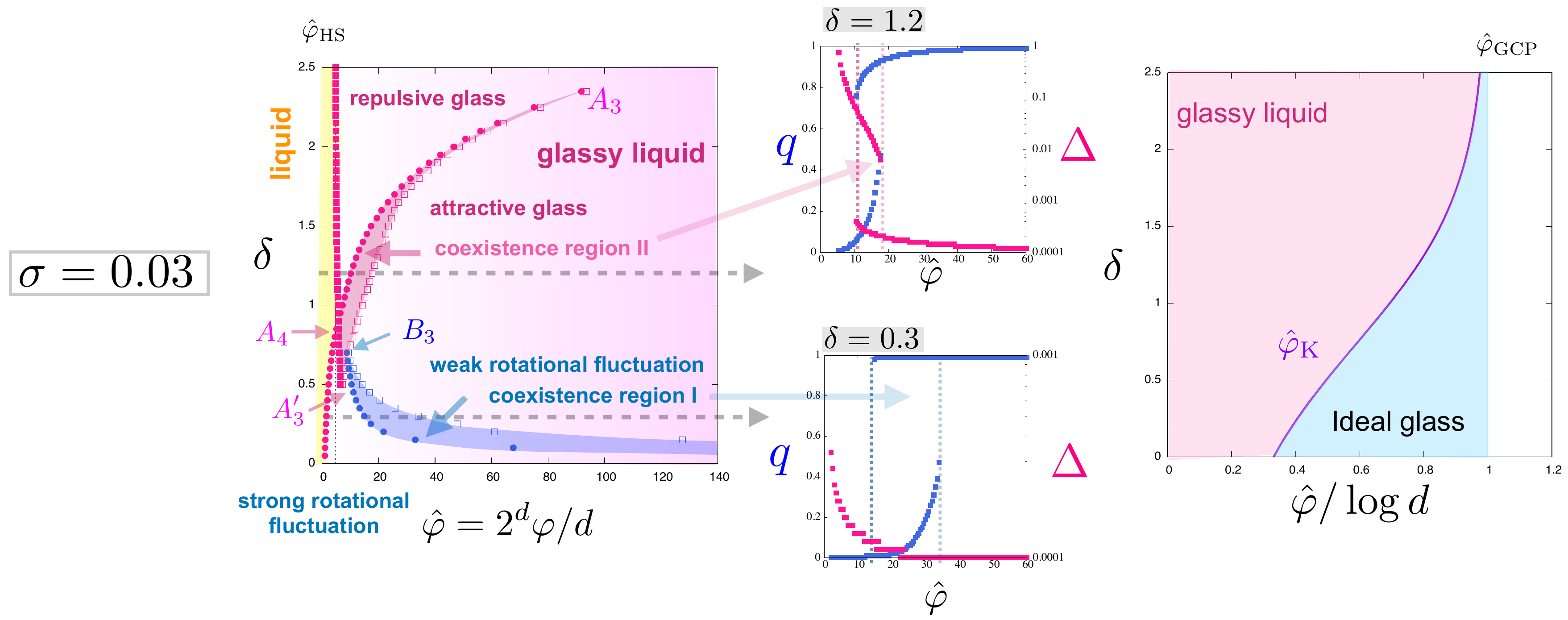}
  \caption{
     (Top panels) Thick patches case: here the width of the attractive well is chosen to be relatively large $\sigma=0.20$.　
     (Bottom panels) Thin patches case: here $\sigma=0.03$.　
     (Left panels)
     Phase diagrams in the liquid and glassy liquid regimes.
     Red filled-squares represent the dynamic glass transition line $\hat\varphi_{\rm d}(\delta)$
        Blue filled-circles/empty-squares represent stability limits
        of weak/strong orientational fluctuations.
        Red filled-circles/empty-squares represent the stability limits
    of attractive/repulsive glasses.
    (Center panels) Behavior of glass order parameters.  (Right panels) Phase diagrams at higher densities.
    The head-tail ratio is fixed as $\delta_{+}/\delta_{-}=0.5$ so that the head-tail symmetry is broken
    and the temperature is fixed at $\hat{T}=0.30$.
     The coverage of the surface by the patches increase with
    $\delta=\delta_{-}$ (vertical axis in the left and right panels).
}
  \label{fig_two_patch_result}
\end{figure*}

\paragraph*{Patchy colloid --}

Let us now we apply the formalism developed above to study patchy colloids with axial symmetry.
In Fig. \ref{fig_patchy_colloid} a) we show the case of two patches
at its head(+) and tail(-). We parameterize the size of the patches 
via $\delta_{\pm}$ which are related to the viewing angles $\theta_{\pm}$ (see
Fig. \ref{fig_patchy_colloid} a)) as
$\delta_{\pm}= \sqrt{d}\cos(\theta_{\pm}/2)$.
We employ the Kern-Frenkel potential  \cite{kern2003fluid}
for the patchy colloids: the particles interact with each other attractively
if their patches {\it touch} but otherwise behave as hardspheres.
The patches {\it touch} if i) the distance $r$ between the centers of the particles
lies inside the potential well of the square-well potential shown in Fig. \ref{fig_patchy_colloid} b)
,i.~e. $D < r < D+\sigma$, and ii) the vector ${\bf r}$ connecting the two centers penetrates the two patches.
Thus the interaction potential can be written as,
\begin{eqnarray}
e^{-\beta V(\xi,x,x')} &=&\theta(\xi)\nonumber\\
& +&(1-e^{1/\hat{T}}) [\theta(\xi-\hat{\sigma})-\theta(\xi)]\Omega(x,x'). \qquad
\label{eq-patch-interaction}
\end{eqnarray}
where $\theta(x)$ is the Heaviside step function.
Note that the potential depends not only on $\xi=d(r/D-1)$ which is a
reduced distance between the centers of the colloids
but also on $x=\hat{r}\cdot {\bf S}_{1}$ and $x'=\hat{r}\cdot {\bf S}_{2}$ where $\hat{r}={\bf r}/r$ is the unit vector parallel to the vector ${\bf r}$
connecting the two centers.
We also introduced a reduced temperature $\hat{T}=k_{\rm B}T/U_{0}$ 
where $U_{0}$ is the depth of the square-well potential shown in Fig.~\ref{fig_patchy_colloid} b) and reduced width of the square-well of the potential $\hat{\sigma}=\sigma/D$.
The function $\Omega(x,x')$ is $1$ if the condition ii) mentioned above is
met and $0$ otherwise.
For the case of the two patches we find,
\beq
\Omega(x,x')=(\theta(-x-\delta_{+})+\theta(x-\delta_{-}))(\theta(x'-\delta_{+})+\theta(-x'-\delta_{-}))\eeq

Let us note that usual sticky colloid is recovered by $\delta_{\pm} = 0$
where our theory becomes the same as the theory by Sellitto and Zamponi \cite{sellitto2013thermodynamic}.

\begin{figure}[h]
 \includegraphics[width=0.4\textwidth]{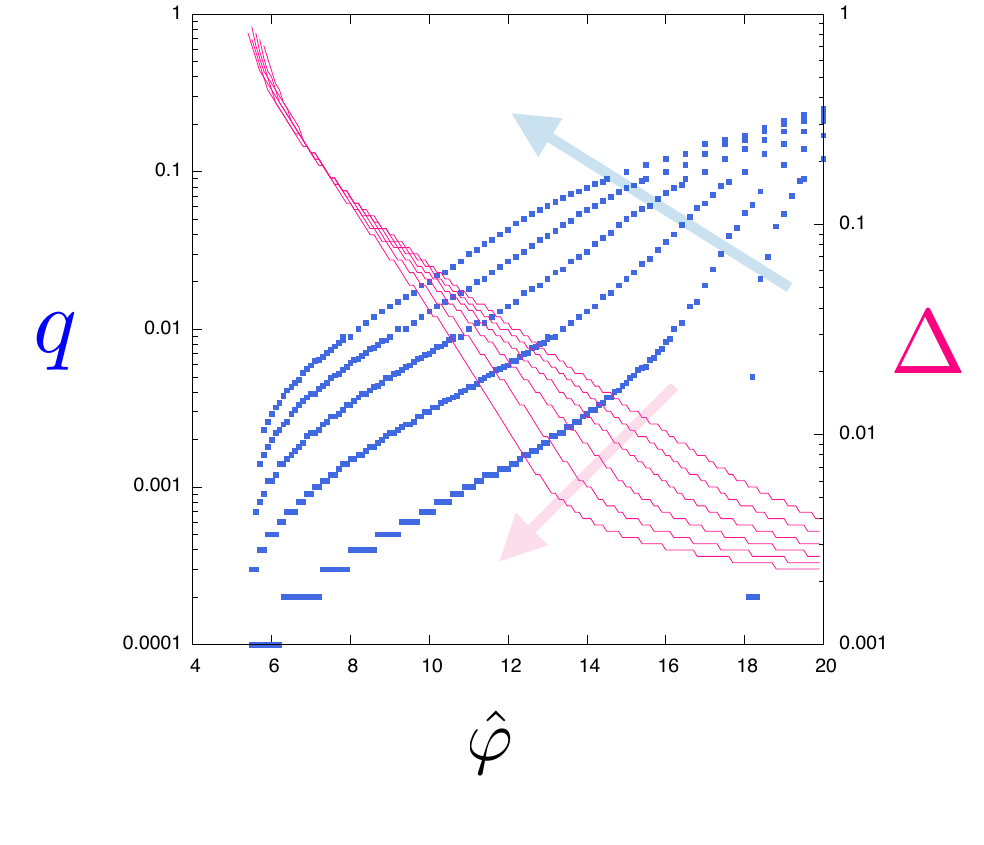}
  \caption{
   The behavior of the glass order parameters with/without head-tail symmetry.
      The head-tail ratio decreases as $\delta_{+}/\delta_{-}=1.0,0.9,0.8,0.7,0.6,0.5$  in the direction indicated by the arrows.
      Here $\hat{T}=0.5$,$\sigma=0.1$ and $\delta_{-}=0.8$. The translational/orientational glass transitions become decoupled only
      in the head-tail symmetric case.
 }
  \label{fig_head_tail_anisotropy}
\end{figure}

In the following we call the ratio $\delta_{+}/\delta_{-}$ as the head-tail ratio and denote $\delta_{-}$ as $\delta$.
The coverage of the surface by the patches increase with $\delta=\delta_{-}$. The patches fully cover the surfaces in the limit $\delta \to 0$
and disappear in $\delta \to \infty$. Thus the system becomes the usual sticky colloid system in $\delta \to 0$ and simple hardsphere system in $\delta \to \infty$.

\begin{figure}[h]
  \includegraphics[width=0.4\textwidth]{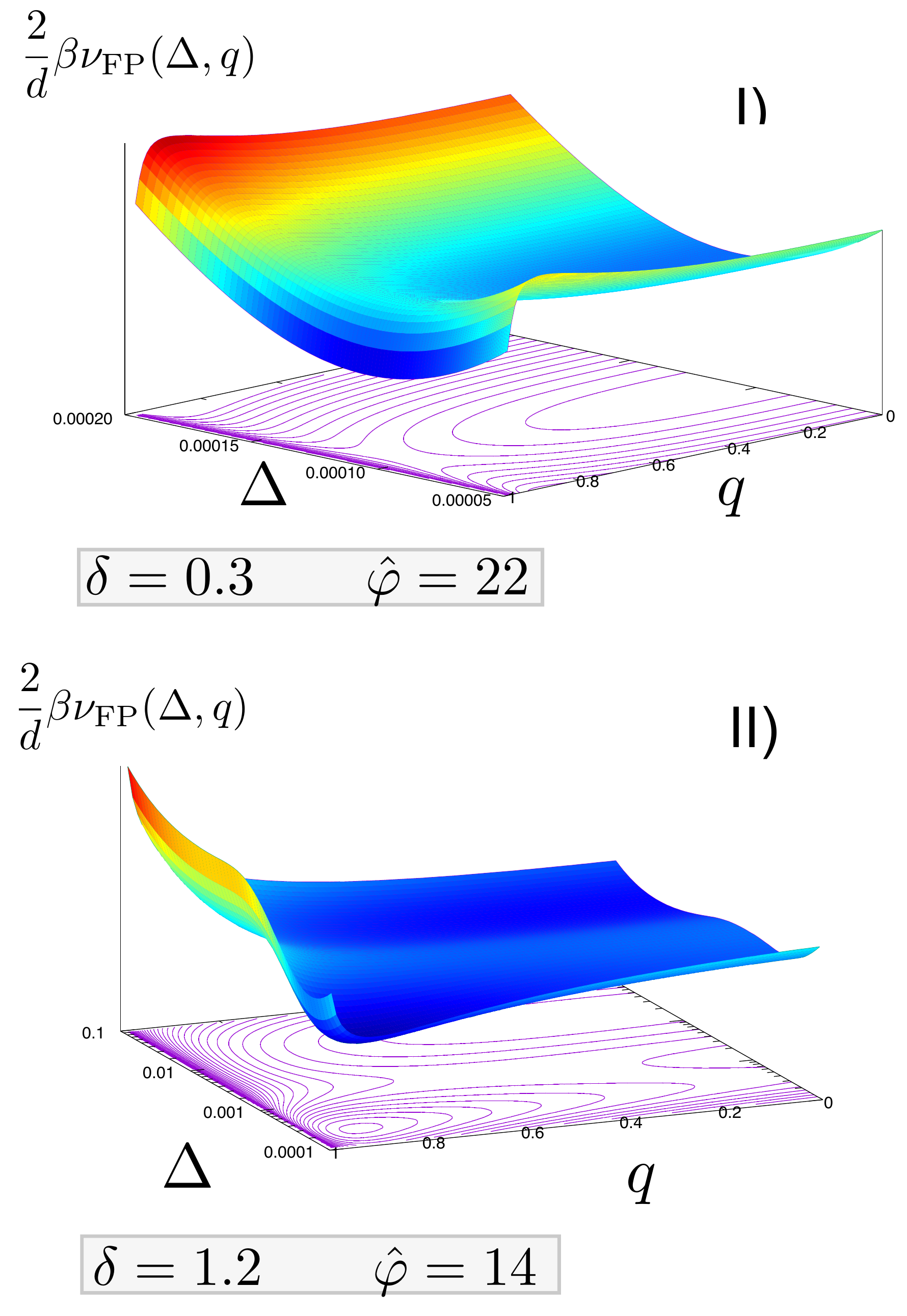}
  \caption{
    Topographic view of the Franz-Parisi Potential with distinct minima.
    Here we show the potential $\nu_{\rm FP}(\Delta,q)$ at representative points located inside the
    coexistence regions I and II of thin-patch system shown in Fig.~\ref{fig_two_patch_result} (bottom)
    in panel I) and II) respectively.
}
  \label{fig_FP_potential}
\end{figure}

\paragraph*{Emergence of glassy liquid}

First let us examine how the glassy states emerge increasing the density $\hat\varphi$. This is done by looking for minima of the Franz-Parisi potential \eq{eq-def-FP}. In the left panels of Fig.~\ref{fig_two_patch_result} we show the phase diagrams of representative cases of relatively thick ($\sigma=0.20$) and thin ($\sigma=0.03$) attractive potential well.

The phase boundary between the liquid and glassy liquid regimes
is given by the dynamical transition density $\varphi_{\rm d}(\delta)$.
We generically find that the dynamical transitions happen simultaneously both in the translational
and orientational degrees of freedom: the solution with $\Delta < \infty$ and $q > 0$ emerge
passing $\varphi_{\rm d}(\delta)$ as can be seen in
Fig.~\ref{fig_two_patch_result} (central panels).
 The exception is the head-tail symmetric case,
i.~e. $\delta_{+}/\delta_{-}=1$ for which the dynamical transitions of the translational and orientational
can decouple (See Fig.~\ref{fig_head_tail_anisotropy}). Such a decoupling has been found by the mode coupling theory and simulations
in hard ellipsoids \cite{letz2000ideal,de2007dynamics}
and hard dumbbells \cite{chong2002idealized,chong2005evidence}.

By construction $\hat\varphi_{\rm d}(\delta)$
converges to that of the simple hardspheres
$\hat\varphi_{\rm HS} \simeq 4.807$  \cite{parisi2010mean} in the limit $\delta \to \infty$ and the usual sticky colloid
in the other limit $\delta \to 0$ 
so that it is natural that the system becomes glassy at a lower density
$\hat\varphi_{\rm d}(0) <  \hat\varphi_{\rm HS}$.
However it can be noticed in the figures that the phase boundary $\varphi_{\rm d}(\delta)$
has a non-monotonic, re-entrant behavior with respect to $\delta$.
Such behaviors are known
in sticky colloids with decreasing temperature  \cite{bergenholtz1999nonergodicity,fabbian1999ideal,
pham2002multiple,eckert2002re,sellitto2013thermodynamic}.
The similarity is reasonable because both the increase of patch size and the lowering
of the temperature enhance the effect of the attractive interaction.

\paragraph*{Kauzmann transition and glass close packing}

Next let us jump to very high density region $\hat\varphi \sim O(\log d)$.
The Kauzmann transition transition takes place in this regime
and $\hat\varphi_{\rm K}$ can be located
as the point where the complexity $\Sigma(1)$ vanishes  \cite{Note1}.
We find,
\beq
(\varphi_{\rm K}/\ln d)^{-1} \simeq 1 + 
(e^{\hat{\sigma}}-1)
[1+(1/\hat{T}-1)e^{1/\hat{T}}]\overline{\Omega} \qquad
\eeq
where $\overline{\Omega}= (\Theta(-\delta_{+}/2)+\Theta(-\delta_{-}/2))^{2}$.
Similarly the glass close packing density
$\hat\varphi_{\rm GCP}$ is located
by the condition $\Sigma(m=0)=0$ which yields
$\varphi_{\rm GCP}/\ln d  \simeq 1$.
As shown in Fig.~\ref{fig_two_patch_result} (right panels),
$\phi_{\rm K}(\delta)$ decreases monotonically with increasing patch size.

\paragraph*{Glass-glass transitions}

Now let us turn to the interior of the super-cooled glassy liquid phase at intermediate densities
$\hat\varphi_{\rm d}(\delta) < \hat\varphi < \hat\varphi_{\rm K}(\delta)$.
There we find quite rich phase behaviors as can be seen in Fig.~\ref{fig_two_patch_result}, with 4 different types of glassy states.

Both in the cases of the thick and thin patches shown Fig.~\ref{fig_two_patch_result} we find a glass-glass transition which is driven by
the orientational degrees of freedom in the region where the effect of the attractive interaction is relatively strong.
Presence and coexistence of two distinct glasses can be clearly seen in the Franz-Parisi potential
shown in Fig.~\ref{fig_FP_potential} I). One minimum has a small $q$ meaning large orientational fluctuations
while the other has a large $q$ meaning weak orientational fluctuations. Interestingly they have almost the same value of $\Delta$
meaning they are almost
the same in the sector of translational degrees of freedom.
The blue points in Fig.~\ref{fig_two_patch_result}
indicate the stability limits of those glasses. The two lines of the stability limits apparently converges at a point
which we denote as as $B_{3}$ point because it is analogous but distinct from the the $A_{3}$ point
which we discuss below. To our knowledge this glass-glass transition was unknown before. 

In the case of the thin-patch, we find another glass-glass transition.
It involves a glass with relatively large $\Delta$ and small $q$ meaning larger translational and orientational
fluctuations and another glass which has smaller $\Delta$ and larger $q$ meaning smaller translational and orientational
fluctuations. Coexistence of the two glasses can be seen again in the Franz-Parisi potential as
shown in Fig.~\ref{fig_FP_potential} II). This glass-glass transition can be considered as essentially the
same as the well known repulsive/attractive glass transition of the simple sticky colloids, which was 1st predicted by the mode coupling theory
\cite{bergenholtz1999nonergodicity,fabbian1999ideal}
and subsequently confirmed by experiments
\cite{pham2002multiple,eckert2002re},
simulations \cite{zaccarelli2002confirmation}
and the replicated liquid theory \cite{sellitto2013thermodynamic}.
The magenta points represent the stability limits of the two glasses.
Just as in the simple sticky colloids, we denote the meeting points of the two lines of the stability limits
as $A_{3}$ point and $A_{3}'$ points. Similarly we denote the point
where the two stability limit lines and the dynamical transition line $\hat\varphi_{\rm d}(\delta)$ meet as $A_{4}$ point.

\paragraph*{Conclusions --}
To conclude we developed a 1st principle theory for glassy phases of uniaxial particles with translational and orientational degrees of freedom, which become exact in the large dimensional limit $d \to \infty$. We applied it to colloids with sticky patches at their heads and tails. We found the system exhibits rich phase behaviors, in particular a novel glass-glass transition
which is driven by the orientational degrees of freedom.
While the $d\to \infty$ limit greatly simplifies the liquid theory and allows theoretical progresses, it would also miss many subtleties of real three-dimensional liquids. Thus it is important to examine our theoretical prediction by experiments on real systems and numerical simulations of finite dimensional systems. For instance, it would be interesting to clarify, much as in the case of the $A_{3}$ point  \cite{fabbian1999ideal,dawson2000higher}, whether the relaxational dynamics exhibit anomalous slow dynamics around the $B_{3}$ point, in the sector of the rotational degrees of freedom or not.
Coexistence of distinct glass phases should be manifested as step-wise response of the system under (de)compression or shear.
For the sticky colloid this has been demonstrated experimentally \cite{pham2006yielding} and theoretically  \cite{altieri2018microscopic}.
On the theoretical side, there are numerous possibilities for future studies extending our results:
exploration of various glassy systems with orientational degrees of freedom including dumbbells and ellipsoids
\cite{letz2000ideal,de2007dynamics,chong2002idealized,chong2005evidence,zheng2011glass,mishra2013two,2018arXiv180701975B},
possibilities of further breaking of the replica symmetry  \cite{charbonneau2014fractal,ikeda2017decoupling} and
state following of the glassy states under compression/shear deformations  \cite{RUYZ14,rainone2016following,urbani2017shear,altieri2018microscopic}.

\begin{acknowledgments}
  We thank Atsushi Ikeda, Harukuni Ikeda, Yuliang Jin, Kota Mitsumoto, Kunimasa Miyazaki, Pierfrancesco Urbani and Francesco Zamponi for useful discussions. This work was supported by KAKENHI (No. 25103005  ``Fluctuation \& Structure'' and No. 50335337) from MEXT, Japan.
  \end{acknowledgments}

\bibliographystyle{mioaps}
\bibliography{ref_yoshino}

\begin{thebibliography}{39}
\expandafter\ifx\csname natexlab\endcsname\relax\def\natexlab#1{#1}\fi
\expandafter\ifx\csname bibnamefont\endcsname\relax
  \def\bibnamefont#1{#1}\fi
\expandafter\ifx\csname bibfnamefont\endcsname\relax
  \def\bibfnamefont#1{#1}\fi
\expandafter\ifx\csname citenamefont\endcsname\relax
  \def\citenamefont#1{#1}\fi
\expandafter\ifx\csname url\endcsname\relax
  \def\url#1{\texttt{#1}}\fi
\expandafter\ifx\csname urlprefix\endcsname\relax\def\urlprefix{URL }\fi
\providecommand{\bibinfo}[2]{#2}
\providecommand{\eprint}[2][]{\url{#2}}

\bibitem[{\citenamefont{Parisi and Zamponi}(2010)}]{parisi2010mean}
\bibinfo{author}{\bibfnamefont{G.}~\bibnamefont{Parisi}} \bibnamefont{and}
  \bibinfo{author}{\bibfnamefont{F.}~\bibnamefont{Zamponi}},
  \bibinfo{journal}{Reviews of Modern Physics} \textbf{\bibinfo{volume}{82}},
  \bibinfo{pages}{789} (\bibinfo{year}{2010}).

\bibitem[{\citenamefont{Kurchan et~al.}(2012)\citenamefont{Kurchan, Parisi, and
  Zamponi}}]{kurchan2012exact}
\bibinfo{author}{\bibfnamefont{J.}~\bibnamefont{Kurchan}},
  \bibinfo{author}{\bibfnamefont{G.}~\bibnamefont{Parisi}}, \bibnamefont{and}
  \bibinfo{author}{\bibfnamefont{F.}~\bibnamefont{Zamponi}},
  \bibinfo{journal}{Journal of Statistical Mechanics: Theory and Experiment}
  \textbf{\bibinfo{volume}{2012}}, \bibinfo{pages}{P10012}
  (\bibinfo{year}{2012}).

\bibitem[{\citenamefont{Kurchan et~al.}(2013)\citenamefont{Kurchan, Parisi,
  Urbani, and Zampoi}}]{kurchan2013exact}
\bibinfo{author}{\bibfnamefont{J.}~\bibnamefont{Kurchan}},
  \bibinfo{author}{\bibfnamefont{G.}~\bibnamefont{Parisi}},
  \bibinfo{author}{\bibfnamefont{P.}~\bibnamefont{Urbani}}, \bibnamefont{and}
  \bibinfo{author}{\bibfnamefont{F.}~\bibnamefont{Zampoi}},
  \bibinfo{journal}{The Journal of Physical Chemistry B}
  \textbf{\bibinfo{volume}{117}}, \bibinfo{pages}{12979}
  (\bibinfo{year}{2013}).

\bibitem[{\citenamefont{Charbonneau
  et~al.}(2014{\natexlab{a}})\citenamefont{Charbonneau, Kurchan, Parisi,
  Urbani, and Zamponi}}]{charbonneau2014exact}
\bibinfo{author}{\bibfnamefont{P.}~\bibnamefont{Charbonneau}},
  \bibinfo{author}{\bibfnamefont{J.}~\bibnamefont{Kurchan}},
  \bibinfo{author}{\bibfnamefont{G.}~\bibnamefont{Parisi}},
  \bibinfo{author}{\bibfnamefont{P.}~\bibnamefont{Urbani}}, \bibnamefont{and}
  \bibinfo{author}{\bibfnamefont{F.}~\bibnamefont{Zamponi}},
  \bibinfo{journal}{Journal of Statistical Mechanics: Theory and Experiment}
  \textbf{\bibinfo{volume}{2014}}, \bibinfo{pages}{P10009}
  (\bibinfo{year}{2014}{\natexlab{a}}).

\bibitem[{\citenamefont{Charbonneau
  et~al.}(2014{\natexlab{b}})\citenamefont{Charbonneau, Kurchan, Parisi,
  Urbani, and Zamponi}}]{charbonneau2014fractal}
\bibinfo{author}{\bibfnamefont{P.}~\bibnamefont{Charbonneau}},
  \bibinfo{author}{\bibfnamefont{J.}~\bibnamefont{Kurchan}},
  \bibinfo{author}{\bibfnamefont{G.}~\bibnamefont{Parisi}},
  \bibinfo{author}{\bibfnamefont{P.}~\bibnamefont{Urbani}}, \bibnamefont{and}
  \bibinfo{author}{\bibfnamefont{F.}~\bibnamefont{Zamponi}},
  \bibinfo{journal}{Nature communications} \textbf{\bibinfo{volume}{5}},
  \bibinfo{pages}{3725} (\bibinfo{year}{2014}{\natexlab{b}}).

\bibitem[{\citenamefont{Suga and Seki}(1974)}]{suga1974thermodynamic}
\bibinfo{author}{\bibfnamefont{H.}~\bibnamefont{Suga}} \bibnamefont{and}
  \bibinfo{author}{\bibfnamefont{S.}~\bibnamefont{Seki}},
  \bibinfo{journal}{Journal of Non-Crystalline Solids}
  \textbf{\bibinfo{volume}{16}}, \bibinfo{pages}{171} (\bibinfo{year}{1974}).

\bibitem[{\citenamefont{Letz et~al.}(2000)\citenamefont{Letz, Schilling, and
  Latz}}]{letz2000ideal}
\bibinfo{author}{\bibfnamefont{M.}~\bibnamefont{Letz}},
  \bibinfo{author}{\bibfnamefont{R.}~\bibnamefont{Schilling}},
  \bibnamefont{and} \bibinfo{author}{\bibfnamefont{A.}~\bibnamefont{Latz}},
  \bibinfo{journal}{Physical Review E} \textbf{\bibinfo{volume}{62}},
  \bibinfo{pages}{5173} (\bibinfo{year}{2000}).

\bibitem[{\citenamefont{Chong and G{\"o}tze}(2002)}]{chong2002idealized}
\bibinfo{author}{\bibfnamefont{S.-H.} \bibnamefont{Chong}} \bibnamefont{and}
  \bibinfo{author}{\bibfnamefont{W.}~\bibnamefont{G{\"o}tze}},
  \bibinfo{journal}{Physical Review E} \textbf{\bibinfo{volume}{65}},
  \bibinfo{pages}{041503} (\bibinfo{year}{2002}).

\bibitem[{\citenamefont{Chong et~al.}(2005)\citenamefont{Chong, Moreno,
  Sciortino, and Kob}}]{chong2005evidence}
\bibinfo{author}{\bibfnamefont{S.-H.} \bibnamefont{Chong}},
  \bibinfo{author}{\bibfnamefont{A.~J.} \bibnamefont{Moreno}},
  \bibinfo{author}{\bibfnamefont{F.}~\bibnamefont{Sciortino}},
  \bibnamefont{and} \bibinfo{author}{\bibfnamefont{W.}~\bibnamefont{Kob}},
  \bibinfo{journal}{Physical review letters} \textbf{\bibinfo{volume}{94}},
  \bibinfo{pages}{215701} (\bibinfo{year}{2005}).

\bibitem[{\citenamefont{De~Michele et~al.}(2007)\citenamefont{De~Michele,
  Schilling, and Sciortino}}]{de2007dynamics}
\bibinfo{author}{\bibfnamefont{C.}~\bibnamefont{De~Michele}},
  \bibinfo{author}{\bibfnamefont{R.}~\bibnamefont{Schilling}},
  \bibnamefont{and}
  \bibinfo{author}{\bibfnamefont{F.}~\bibnamefont{Sciortino}},
  \bibinfo{journal}{Physical review letters} \textbf{\bibinfo{volume}{98}},
  \bibinfo{pages}{265702} (\bibinfo{year}{2007}).

\bibitem[{\citenamefont{Zheng et~al.}(2011)\citenamefont{Zheng, Wang, Han
  et~al.}}]{zheng2011glass}
\bibinfo{author}{\bibfnamefont{Z.}~\bibnamefont{Zheng}},
  \bibinfo{author}{\bibfnamefont{F.}~\bibnamefont{Wang}},
  \bibinfo{author}{\bibfnamefont{Y.}~\bibnamefont{Han}}, \bibnamefont{et~al.},
  \bibinfo{journal}{Physical review letters} \textbf{\bibinfo{volume}{107}},
  \bibinfo{pages}{065702} (\bibinfo{year}{2011}).

\bibitem[{\citenamefont{Mishra et~al.}(2013)\citenamefont{Mishra, Rangarajan,
  and Ganapathy}}]{mishra2013two}
\bibinfo{author}{\bibfnamefont{C.~K.} \bibnamefont{Mishra}},
  \bibinfo{author}{\bibfnamefont{A.}~\bibnamefont{Rangarajan}},
  \bibnamefont{and}
  \bibinfo{author}{\bibfnamefont{R.}~\bibnamefont{Ganapathy}},
  \bibinfo{journal}{Physical review letters} \textbf{\bibinfo{volume}{110}},
  \bibinfo{pages}{188301} (\bibinfo{year}{2013}).

\bibitem[{\citenamefont{M{\'e}zard and Parisi}(1999)}]{mezard1999first}
\bibinfo{author}{\bibfnamefont{M.}~\bibnamefont{M{\'e}zard}} \bibnamefont{and}
  \bibinfo{author}{\bibfnamefont{G.}~\bibnamefont{Parisi}},
  \bibinfo{journal}{The Journal of chemical physics}
  \textbf{\bibinfo{volume}{111}}, \bibinfo{pages}{1076} (\bibinfo{year}{1999}).

\bibitem[{\citenamefont{Yoshino}(2018)}]{yoshino2018}
\bibinfo{author}{\bibfnamefont{H.}~\bibnamefont{Yoshino}},
  \bibinfo{journal}{SciPost Phys.} \textbf{\bibinfo{volume}{4}},
  \bibinfo{pages}{40} (\bibinfo{year}{2018}).

\bibitem[{\citenamefont{Chen et~al.}(2011)\citenamefont{Chen, Bae, and
  Granick}}]{chen2011directed}
\bibinfo{author}{\bibfnamefont{Q.}~\bibnamefont{Chen}},
  \bibinfo{author}{\bibfnamefont{S.~C.} \bibnamefont{Bae}}, \bibnamefont{and}
  \bibinfo{author}{\bibfnamefont{S.}~\bibnamefont{Granick}},
  \bibinfo{journal}{Nature} \textbf{\bibinfo{volume}{469}},
  \bibinfo{pages}{381} (\bibinfo{year}{2011}).

\bibitem[{\citenamefont{Sciortino}(2008)}]{sciortino2008gel}
\bibinfo{author}{\bibfnamefont{F.}~\bibnamefont{Sciortino}},
  \bibinfo{journal}{The European Physical Journal B}
  \textbf{\bibinfo{volume}{64}}, \bibinfo{pages}{505} (\bibinfo{year}{2008}).

\bibitem[{\citenamefont{Bianchi et~al.}(2006)\citenamefont{Bianchi, Largo,
  Tartaglia, Zaccarelli, and Sciortino}}]{bianchi2006phase}
\bibinfo{author}{\bibfnamefont{E.}~\bibnamefont{Bianchi}},
  \bibinfo{author}{\bibfnamefont{J.}~\bibnamefont{Largo}},
  \bibinfo{author}{\bibfnamefont{P.}~\bibnamefont{Tartaglia}},
  \bibinfo{author}{\bibfnamefont{E.}~\bibnamefont{Zaccarelli}},
  \bibnamefont{and}
  \bibinfo{author}{\bibfnamefont{F.}~\bibnamefont{Sciortino}},
  \bibinfo{journal}{Physical review letters} \textbf{\bibinfo{volume}{97}},
  \bibinfo{pages}{168301} (\bibinfo{year}{2006}).

\bibitem[{\citenamefont{Ruzicka et~al.}(2011)\citenamefont{Ruzicka, Zaccarelli,
  Zulian, Angelini, Sztucki, Moussa{\"\i}d, Narayanan, and
  Sciortino}}]{ruzicka2011observation}
\bibinfo{author}{\bibfnamefont{B.}~\bibnamefont{Ruzicka}},
  \bibinfo{author}{\bibfnamefont{E.}~\bibnamefont{Zaccarelli}},
  \bibinfo{author}{\bibfnamefont{L.}~\bibnamefont{Zulian}},
  \bibinfo{author}{\bibfnamefont{R.}~\bibnamefont{Angelini}},
  \bibinfo{author}{\bibfnamefont{M.}~\bibnamefont{Sztucki}},
  \bibinfo{author}{\bibfnamefont{A.}~\bibnamefont{Moussa{\"\i}d}},
  \bibinfo{author}{\bibfnamefont{T.}~\bibnamefont{Narayanan}},
  \bibnamefont{and}
  \bibinfo{author}{\bibfnamefont{F.}~\bibnamefont{Sciortino}},
  \bibinfo{journal}{Nature materials} \textbf{\bibinfo{volume}{10}},
  \bibinfo{pages}{56} (\bibinfo{year}{2011}).

\bibitem[{\citenamefont{Bergenholtz and
  Fuchs}(1999)}]{bergenholtz1999nonergodicity}
\bibinfo{author}{\bibfnamefont{J.}~\bibnamefont{Bergenholtz}} \bibnamefont{and}
  \bibinfo{author}{\bibfnamefont{M.}~\bibnamefont{Fuchs}},
  \bibinfo{journal}{Physical Review E} \textbf{\bibinfo{volume}{59}},
  \bibinfo{pages}{5706} (\bibinfo{year}{1999}).

\bibitem[{\citenamefont{Fabbian et~al.}(1999)\citenamefont{Fabbian, G{\"o}tze,
  Sciortino, Tartaglia, and Thiery}}]{fabbian1999ideal}
\bibinfo{author}{\bibfnamefont{L.}~\bibnamefont{Fabbian}},
  \bibinfo{author}{\bibfnamefont{W.}~\bibnamefont{G{\"o}tze}},
  \bibinfo{author}{\bibfnamefont{F.}~\bibnamefont{Sciortino}},
  \bibinfo{author}{\bibfnamefont{P.}~\bibnamefont{Tartaglia}},
  \bibnamefont{and} \bibinfo{author}{\bibfnamefont{F.}~\bibnamefont{Thiery}},
  \bibinfo{journal}{Physical Review E} \textbf{\bibinfo{volume}{59}},
  \bibinfo{pages}{R1347} (\bibinfo{year}{1999}).

\bibitem[{\citenamefont{Pham et~al.}(2002)\citenamefont{Pham, Puertas,
  Bergenholtz, Egelhaaf, Moussa{\i}d, Pusey, Schofield, Cates, Fuchs, and
  Poon}}]{pham2002multiple}
\bibinfo{author}{\bibfnamefont{K.~N.} \bibnamefont{Pham}},
  \bibinfo{author}{\bibfnamefont{A.~M.} \bibnamefont{Puertas}},
  \bibinfo{author}{\bibfnamefont{J.}~\bibnamefont{Bergenholtz}},
  \bibinfo{author}{\bibfnamefont{S.~U.} \bibnamefont{Egelhaaf}},
  \bibinfo{author}{\bibfnamefont{A.}~\bibnamefont{Moussa{\i}d}},
  \bibinfo{author}{\bibfnamefont{P.~N.} \bibnamefont{Pusey}},
  \bibinfo{author}{\bibfnamefont{A.~B.} \bibnamefont{Schofield}},
  \bibinfo{author}{\bibfnamefont{M.~E.} \bibnamefont{Cates}},
  \bibinfo{author}{\bibfnamefont{M.}~\bibnamefont{Fuchs}}, \bibnamefont{and}
  \bibinfo{author}{\bibfnamefont{W.~C.} \bibnamefont{Poon}},
  \bibinfo{journal}{Science} \textbf{\bibinfo{volume}{296}},
  \bibinfo{pages}{104} (\bibinfo{year}{2002}).

\bibitem[{\citenamefont{Eckert and Bartsch}(2002)}]{eckert2002re}
\bibinfo{author}{\bibfnamefont{T.}~\bibnamefont{Eckert}} \bibnamefont{and}
  \bibinfo{author}{\bibfnamefont{E.}~\bibnamefont{Bartsch}},
  \bibinfo{journal}{Physical review letters} \textbf{\bibinfo{volume}{89}},
  \bibinfo{pages}{125701} (\bibinfo{year}{2002}).

\bibitem[{\citenamefont{Pham et~al.}(2006)\citenamefont{Pham, Petekidis,
  Vlassopoulos, Egelhaaf, Pusey, and Poon}}]{pham2006yielding}
\bibinfo{author}{\bibfnamefont{K.}~\bibnamefont{Pham}},
  \bibinfo{author}{\bibfnamefont{G.}~\bibnamefont{Petekidis}},
  \bibinfo{author}{\bibfnamefont{D.}~\bibnamefont{Vlassopoulos}},
  \bibinfo{author}{\bibfnamefont{S.}~\bibnamefont{Egelhaaf}},
  \bibinfo{author}{\bibfnamefont{P.}~\bibnamefont{Pusey}}, \bibnamefont{and}
  \bibinfo{author}{\bibfnamefont{W.}~\bibnamefont{Poon}}, \bibinfo{journal}{EPL
  (Europhysics Letters)} \textbf{\bibinfo{volume}{75}}, \bibinfo{pages}{624}
  (\bibinfo{year}{2006}).

\bibitem[{\citenamefont{Sciortino}(2002)}]{sciortino2002disordered}
\bibinfo{author}{\bibfnamefont{F.}~\bibnamefont{Sciortino}},
  \bibinfo{journal}{Nature materials} \textbf{\bibinfo{volume}{1}},
  \bibinfo{pages}{145} (\bibinfo{year}{2002}).

\bibitem[{\citenamefont{Sellitto and
  Zamponi}(2013)}]{sellitto2013thermodynamic}
\bibinfo{author}{\bibfnamefont{M.}~\bibnamefont{Sellitto}} \bibnamefont{and}
  \bibinfo{author}{\bibfnamefont{F.}~\bibnamefont{Zamponi}},
  \bibinfo{journal}{EPL (Europhysics Letters)} \textbf{\bibinfo{volume}{103}},
  \bibinfo{pages}{46005} (\bibinfo{year}{2013}).

\bibitem[{Note1()}]{Note1}
Note1, \bibinfo{note}{supplemental Material}.

\bibitem[{\citenamefont{Monasson}(1995)}]{Mo95}
\bibinfo{author}{\bibfnamefont{R.}~\bibnamefont{Monasson}},
  \bibinfo{journal}{Phys. Rev. Lett.} \textbf{\bibinfo{volume}{75}},
  \bibinfo{pages}{2847} (\bibinfo{year}{1995}).

\bibitem[{\citenamefont{Kern and Frenkel}(2003)}]{kern2003fluid}
\bibinfo{author}{\bibfnamefont{N.}~\bibnamefont{Kern}} \bibnamefont{and}
  \bibinfo{author}{\bibfnamefont{D.}~\bibnamefont{Frenkel}},
  \bibinfo{journal}{The Journal of chemical physics}
  \textbf{\bibinfo{volume}{118}}, \bibinfo{pages}{9882} (\bibinfo{year}{2003}).

\bibitem[{\citenamefont{Zaccarelli et~al.}(2002)\citenamefont{Zaccarelli,
  Foffi, Dawson, Buldyrev, Sciortino, and
  Tartaglia}}]{zaccarelli2002confirmation}
\bibinfo{author}{\bibfnamefont{E.}~\bibnamefont{Zaccarelli}},
  \bibinfo{author}{\bibfnamefont{G.}~\bibnamefont{Foffi}},
  \bibinfo{author}{\bibfnamefont{K.~A.} \bibnamefont{Dawson}},
  \bibinfo{author}{\bibfnamefont{S.}~\bibnamefont{Buldyrev}},
  \bibinfo{author}{\bibfnamefont{F.}~\bibnamefont{Sciortino}},
  \bibnamefont{and}
  \bibinfo{author}{\bibfnamefont{P.}~\bibnamefont{Tartaglia}},
  \bibinfo{journal}{Physical Review E} \textbf{\bibinfo{volume}{66}},
  \bibinfo{pages}{041402} (\bibinfo{year}{2002}).

\bibitem[{\citenamefont{Dawson et~al.}(2000)\citenamefont{Dawson, Foffi, Fuchs,
  G{\"o}tze, Sciortino, Sperl, Tartaglia, Voigtmann, and
  Zaccarelli}}]{dawson2000higher}
\bibinfo{author}{\bibfnamefont{K.}~\bibnamefont{Dawson}},
  \bibinfo{author}{\bibfnamefont{G.}~\bibnamefont{Foffi}},
  \bibinfo{author}{\bibfnamefont{M.}~\bibnamefont{Fuchs}},
  \bibinfo{author}{\bibfnamefont{W.}~\bibnamefont{G{\"o}tze}},
  \bibinfo{author}{\bibfnamefont{F.}~\bibnamefont{Sciortino}},
  \bibinfo{author}{\bibfnamefont{M.}~\bibnamefont{Sperl}},
  \bibinfo{author}{\bibfnamefont{P.}~\bibnamefont{Tartaglia}},
  \bibinfo{author}{\bibfnamefont{T.}~\bibnamefont{Voigtmann}},
  \bibnamefont{and}
  \bibinfo{author}{\bibfnamefont{E.}~\bibnamefont{Zaccarelli}},
  \bibinfo{journal}{Physical Review E} \textbf{\bibinfo{volume}{63}},
  \bibinfo{pages}{011401} (\bibinfo{year}{2000}).

\bibitem[{\citenamefont{{Altieri} et~al.}(2018)\citenamefont{{Altieri},
  {Urbani}, and {Zamponi}}}]{altieri2018microscopic}
\bibinfo{author}{\bibfnamefont{A.}~\bibnamefont{{Altieri}}},
  \bibinfo{author}{\bibfnamefont{P.}~\bibnamefont{{Urbani}}}, \bibnamefont{and}
  \bibinfo{author}{\bibfnamefont{F.}~\bibnamefont{{Zamponi}}},
  \bibinfo{journal}{ArXiv e-prints}  (\bibinfo{year}{2018}),
  \eprint{1806.05453}.

\bibitem[{\citenamefont{{Brito} et~al.}(2018)\citenamefont{{Brito}, {Ikeda},
  {Urbani}, {Wyart}, and {Zamponi}}}]{2018arXiv180701975B}
\bibinfo{author}{\bibfnamefont{C.}~\bibnamefont{{Brito}}},
  \bibinfo{author}{\bibfnamefont{H.}~\bibnamefont{{Ikeda}}},
  \bibinfo{author}{\bibfnamefont{P.}~\bibnamefont{{Urbani}}},
  \bibinfo{author}{\bibfnamefont{M.}~\bibnamefont{{Wyart}}}, \bibnamefont{and}
  \bibinfo{author}{\bibfnamefont{F.}~\bibnamefont{{Zamponi}}},
  \bibinfo{journal}{ArXiv e-prints}  (\bibinfo{year}{2018}),
  \eprint{1807.01975}.

\bibitem[{\citenamefont{{Ikeda} et~al.}(2017)\citenamefont{{Ikeda}, {Miyazaki},
  {Yoshino}, and {Ikeda}}}]{ikeda2017decoupling}
\bibinfo{author}{\bibfnamefont{H.}~\bibnamefont{{Ikeda}}},
  \bibinfo{author}{\bibfnamefont{K.}~\bibnamefont{{Miyazaki}}},
  \bibinfo{author}{\bibfnamefont{H.}~\bibnamefont{{Yoshino}}},
  \bibnamefont{and} \bibinfo{author}{\bibfnamefont{A.}~\bibnamefont{{Ikeda}}},
  \bibinfo{journal}{ArXiv e-prints}  (\bibinfo{year}{2017}),
  \eprint{1710.08373}.

\bibitem[{\citenamefont{Rainone et~al.}(2015)\citenamefont{Rainone, Urbani,
  Yoshino, and Zamponi}}]{RUYZ14}
\bibinfo{author}{\bibfnamefont{C.}~\bibnamefont{Rainone}},
  \bibinfo{author}{\bibfnamefont{P.}~\bibnamefont{Urbani}},
  \bibinfo{author}{\bibfnamefont{H.}~\bibnamefont{Yoshino}}, \bibnamefont{and}
  \bibinfo{author}{\bibfnamefont{F.}~\bibnamefont{Zamponi}},
  \bibinfo{journal}{Physical review letters} \textbf{\bibinfo{volume}{114}},
  \bibinfo{pages}{015701} (\bibinfo{year}{2015}).

\bibitem[{\citenamefont{Rainone and Urbani}(2016)}]{rainone2016following}
\bibinfo{author}{\bibfnamefont{C.}~\bibnamefont{Rainone}} \bibnamefont{and}
  \bibinfo{author}{\bibfnamefont{P.}~\bibnamefont{Urbani}},
  \bibinfo{journal}{Journal of Statistical Mechanics: Theory and Experiment}
  \textbf{\bibinfo{volume}{2016}}, \bibinfo{pages}{053302}
  (\bibinfo{year}{2016}).

\bibitem[{\citenamefont{Urbani and Zamponi}(2017)}]{urbani2017shear}
\bibinfo{author}{\bibfnamefont{P.}~\bibnamefont{Urbani}} \bibnamefont{and}
  \bibinfo{author}{\bibfnamefont{F.}~\bibnamefont{Zamponi}},
  \bibinfo{journal}{Physical review letters} \textbf{\bibinfo{volume}{118}},
  \bibinfo{pages}{038001} (\bibinfo{year}{2017}).

\bibitem[{\citenamefont{Parisi}(1979)}]{parisi1979infinite}
\bibinfo{author}{\bibfnamefont{G.}~\bibnamefont{Parisi}},
  \bibinfo{journal}{Physical Review Letters} \textbf{\bibinfo{volume}{43}},
  \bibinfo{pages}{1754} (\bibinfo{year}{1979}).

\bibitem[{\citenamefont{Gentle}(2007)}]{james2007gentle}
\bibinfo{author}{\bibfnamefont{J.~E.} \bibnamefont{Gentle}},
  \emph{\bibinfo{title}{Matrix Algebra: Theory, Computations, and Applications
  in Statistics}} (\bibinfo{publisher}{Springer}, \bibinfo{year}{2007}).

\bibitem[{\citenamefont{Duplantier}(1981)}]{duplantier1981comment}
\bibinfo{author}{\bibfnamefont{B.}~\bibnamefont{Duplantier}},
  \bibinfo{journal}{Journal of Physics A: Mathematical and General}
  \textbf{\bibinfo{volume}{14}}, \bibinfo{pages}{283} (\bibinfo{year}{1981}).

\end{thebibliography}

\onecolumngrid
%%\documentclass[aps,prl,amsmath,amsfonts,floatfix,superscriptaddress,nofootinbib,twocolumn]{revtex4-1}
%\documentclass[aps,prl,amsmath,amsfonts,floatfix,superscriptaddress,nofootinbib,notitlepage]{revtex4-1}
%%\documentclass[prl,amsmath,amsfonts,superscriptaddress,nofootinbib,twocolumn]{revtex4-1}
%\usepackage[colorlinks=true]{hyperref}
%\hypersetup{
%  colorlinks=true,
%  linkcolor=blue,
%  filecolor=magenta,
%  citecolor=blue,
%  urlcolor=blue,
%}
%
%\usepackage{graphicx,color}
%\usepackage{bbm}
%%\usepackage{multicol}
%%\usepackage{multirow}
%\usepackage{mathrsfs} \usepackage{amssymb} 
%\usepackage{commath}\usepackage{tikz}
%\usetikzlibrary{shapes,arrows,positioning,calc}\usepackage{stmaryrd}\usepackage{bm}
%
%%\font\msytw=msbm9 scaled\magstep1
%
%\def\vr{\Vec{r}}
%\def\vR{\Vec{R}}
%\def\vu{\Vec{u}}
%\def\veta{\bm \eta}
%\def\vS{\Vec{S}}
%\def\oS{\Vec{\overline{S}}}
%
%\newcommand{\av}[2]{\left\langle #1 \right\rangle_{#2}}
%
%\newcommand{\ov}[1]{\overline{#1}}
%\newcommand{\ovS}{\overline{\Vec{S}}}
%\newcommand{\ovr}{\overline{\Vec{r}}}
%\newcommand{\ovu}{\overline{\Vec{u}}}
%
%\newcommand{\eq}[1]{Eq.~(\ref{#1})}
%\newcommand{\beq}{\begin{equation}} \newcommand{\eeq}{\end{equation}}
%\newcommand{\beqn}{\begin{eqnarray}} \newcommand{\eeqn}{\end{eqnarray}}
%\newcommand{\wh}{\widehat} \newcommand{\wt}{\widetilde}
%\newcommand{\Tr}{\mbox{Tr}}
%
%\newcommand{\cD}{{\cal D}}
%
%\newcommand{\blue}[1]{\textcolor{blue}{#1}}
%\newcommand{\red}[1]{\textcolor{red}{#1}}
%\renewcommand{\Vec}[1]{{\bf #1}}
%
%\newcommand{\intS}{\int_{S}}
%\newcommand{\intr}{\int d^{d}r}
%
%\newcommand{\bs}{\blacksquare}
%
%\begin{document}

\begin{center}
  {\bf
  Supplemental Material for:\\ Translational and orientational glass transitions in the large-dimensional limit : replica theory and application to patchy colloids}
\end{center}

%
%\title{Supplemental Material for:\\ Translational and orientational glass transitions in the large-dimensional limit : replica theory and application to patchy colloids}
%
%\author{Hajime Yoshino}
%\affiliation{Cybermedia Center, Osaka University, Toyonaka, Osaka 560-0043, Japan}
%\affiliation{Graduate School of Science, Osaka University, Toyonaka, Osaka 560-0043, Japan}
%\email[⟨Email:⟩]{yoshino@cmc.osaka-u.ac.jp}
%
%%\date{\today}
%
%\maketitle
%
\nopagebreak

 \section{Assembly of anisotropic particles}

       \subsection{Model}

We consider an assembly of {\it anisotropic} particles with axial symmetry in $d$-dimensional space interacting with each other through a two-body potential,
\beq
H=\sum_{i < j} V(\vr_{ij},\vS_{i},\vS_{j})
\eeq
where $\vr_{i}$ and $\vS_{i}$
($i=1,2,\ldots,N$) are $d$-dimensional vectors
representing the position and orientation of the particles.
%The latter is, for instance, the director of Janus particles.
The specific form of the potential is not important but 
we assume the system is orientationally and translationally
invariant so that it can be parameterized as,
\beq
V(\vr_{12},\vS_{1},\vS_{2})=
V(r_{12},\hat{\bf r}_{12}\cdot \vS_{1},\hat{\bf r}_{12}\cdot \vS_{2},\vS_{1}\cdot\vS_{2})
\label{eq-sm-orientationally-invariant-potential}
\eeq
where $\vr_{12}=\vr_{1}-\vr_{2}$, $\hat{r}=\vr/r$ and $r=|\vr|$.
In the following we set up a mean-field theoretical framework to describe liquid and glass states of such a system in the $d\to \infty$ limit.

We denote the number density of the system as $\rho=N/V$ with $V$ being the volume of the container. For the colloids including the patchy colloids it is useful to introduce
the volume fraction
\beq
\varphi=\rho (\Omega_{d}/d)(D/2)^{d}
\label{eq-sm-vaphi-rho}
\eeq
where $D$ is the diameter of the particle
and $\Omega_{d}$ is the surface area of unit sphere in $d$-dimensions.
We also introduce the reduced volume fraction $\hat\varphi=2^{d}\varphi/d$ 
which is useful to consider $d \to \infty$ limit  \cite{parisi2010mean}.

\section{Density functional theory for anisotropic particles}

Let us introduce a generalized density field,
\beq
\rho(\vr,\vS) \equiv \sum_{i=1}^{N}\delta(\vr-\vr_{i})\delta(\vS-\vS_{i}).
\eeq
which is normalized such that
\beq
\int d^{d}r d\vS \rho(\vr,\vS) =N.
\eeq
By recalling the fact that in the $d \to \infty$ limit so that the free-energy of liquids
can be expressed exactly by 1st virial expansion we find  \cite{kurchan2012exact,yoshino2018},  
\begin{eqnarray}
 -\beta {\cal F}[\rho(\vr,\vS)]=
  - \int_{\vr,\vS} \rho(\vr,\vS)(\ln \rho (\vr,\vS) -1)
  +\frac{1}{2} \int_{\vr_{1},\vS_{1},\vr_{2},\vS_{2}}
  \rho(\vr_{1},\vS_{1})  \rho(\vr_{2},\vS_{2})
  f(\vr_{12},\vS_{1},\vS_{2}) \qquad 
%  \label{eq-sm-free-energy-functional-1replica}
\end{eqnarray}
where we introduced a shorthanded notation $\int_{\vr,\vS}\equiv\intr \intS d\vS$
and the Mayer function,
\beq
f(\vr_{12},\vS_{1},\vS_{2})=
e^{-\beta V(\vr_{12},\vS_{1},\vS_{2})}
\eeq
Here the potential $V(\vr_{12},\vS_{1},\vS_{2})$
is orientationally and translationally
invariant as in  \eq{eq-sm-orientationally-invariant-potential}

\section{Replicated liquid of anisotropic particles}

Next we consider the replicated liquid of the anisotropic particles
made of replicas $a=1,2,\ldots,m$ obeying the Hamiltonian,
\beq
H_{m}=\sum_{a=1}^{m}\sum_{i<j}
V(\vr^{a}_{ij},\vS_{i}^{a},\vS_{j}^{a}).
\eeq
We denote the position of the particles as $\ovr_{i}=(\vr_{i}^{1},\vr_{i}^{2},\ldots,\vr_{i}^{m})$.
%as \eq{eq-sm-def-ovS}
By introducing the replicated density field,
\beq
\rho(\Vec{\overline{r}},\Vec{\overline{S}})=\sum_{i=1}^{N}
\prod_{a=1}^{m}\delta(\vr^{a}-\vr_{i}^{a})\delta(\vS^{a}-\vS_{i}^{a})
\eeq
we can write the free-energy of the replicated system as,
\beq
- N \beta m  \phi_{m}[\rho(\overline{\Vec{r}},\ovS)]=
-\int_{\ovr,\ovS} \rho(\ovr,\ovS) (\ln \rho (\ovr,\ovS) -1)
+\frac{1}{2}\int_{\ovr_{1},\ovS_{1},\ovr_{2},\ovS_{2}}
\rho(\ovr_{1},\ovS_{1})\rho(\ovr_{2},\ovS_{2})
f_{m}(\ovr_{12},\ovS_{1},\ovS_{2})
\label{eq-sm-free-energy-functional-nreplica-anistropic-particle}
\eeq
where we introduced the replicated Mayer function,
\beq
f_{m}(\ovr_{12},\ovS_{1},\ovS_{2})=\prod_{a=1}^{m}
e^{-\beta V(\vr_{12}^{a},\vS_{1}^{a},\vS_{2}^{a})}-1
\eeq
Here the potential $V(\ovr_{12},\ovS_{1},\ovS_{2})$ is orientationally and translationally invariant as in \eq{eq-sm-orientationally-invariant-potential}.

As in the case of simple spheres  \cite{kurchan2012exact} we decompose the spatial coordinate of particles as,
\beq
\vr_{i}^{a}=\vR_{i}+\vu^{a}_{i}
\label{eq-sm-decomposition-particle-coordinate}
\eeq
where
\beq
\vR_{i}=\frac{1}{m}\sum_{a=1}^{m}\vr_{i}^{a}
\label{eq-sm-center-of-mass}
\eeq
is the center of mass position of the molecule made of replicas $a=1,2,\ldots,m$ and $\vu^{a}_{i}$ represents fluctuation within the molecule.
Note that
\beq
\sum_{a=1}^{m} \vu^{a}_{i}=0
\label{eq-sm-sum-rule0}
\eeq
holds by definition.
%Here $D$ represents a characteristic microscopic length scale such as the diameter of
%the particle.
The natural glass order parameters which are invariant
under global translations and rotations of all replicas is,
\beqn
q_{ab} &=&\lim_{N \to \infty}\frac{1}{N}\sum_{i=1}^{N}\langle \vu^{a}_{i} \cdot \vu^{b}_{i} \rangle.  \\
Q_{ab} &=&\lim_{N \to \infty}\frac{1}{Nd}\sum_{i=1}^{N} \langle \vS^{a}_{i}\cdot \vS^{b}_{i}\rangle. \\
P_{ab}&=&\lim_{N \to \infty}\frac{1}{N\sqrt{d}}\sum_{i=1}^{N} \langle \vu^{a}_{i}\cdot \vS^{b}_{i}\rangle.
\label{eq-sm-def-qab}
\eeqn
Note that the following sum rules hold because of the identity \eq{eq-sm-sum-rule0},
\beq
\sum_{b=1}^{m}q_{ab}=\sum_{a=1}^{m}q_{ab}=0 \qquad
\sum_{b=1}^{m}P_{ab}=\sum_{a=1}^{m}P_{ab}=0
\label{eq-sm-sum-rule}
\eeq

\begin{figure}[t]
  \includegraphics[width=0.8\textwidth]{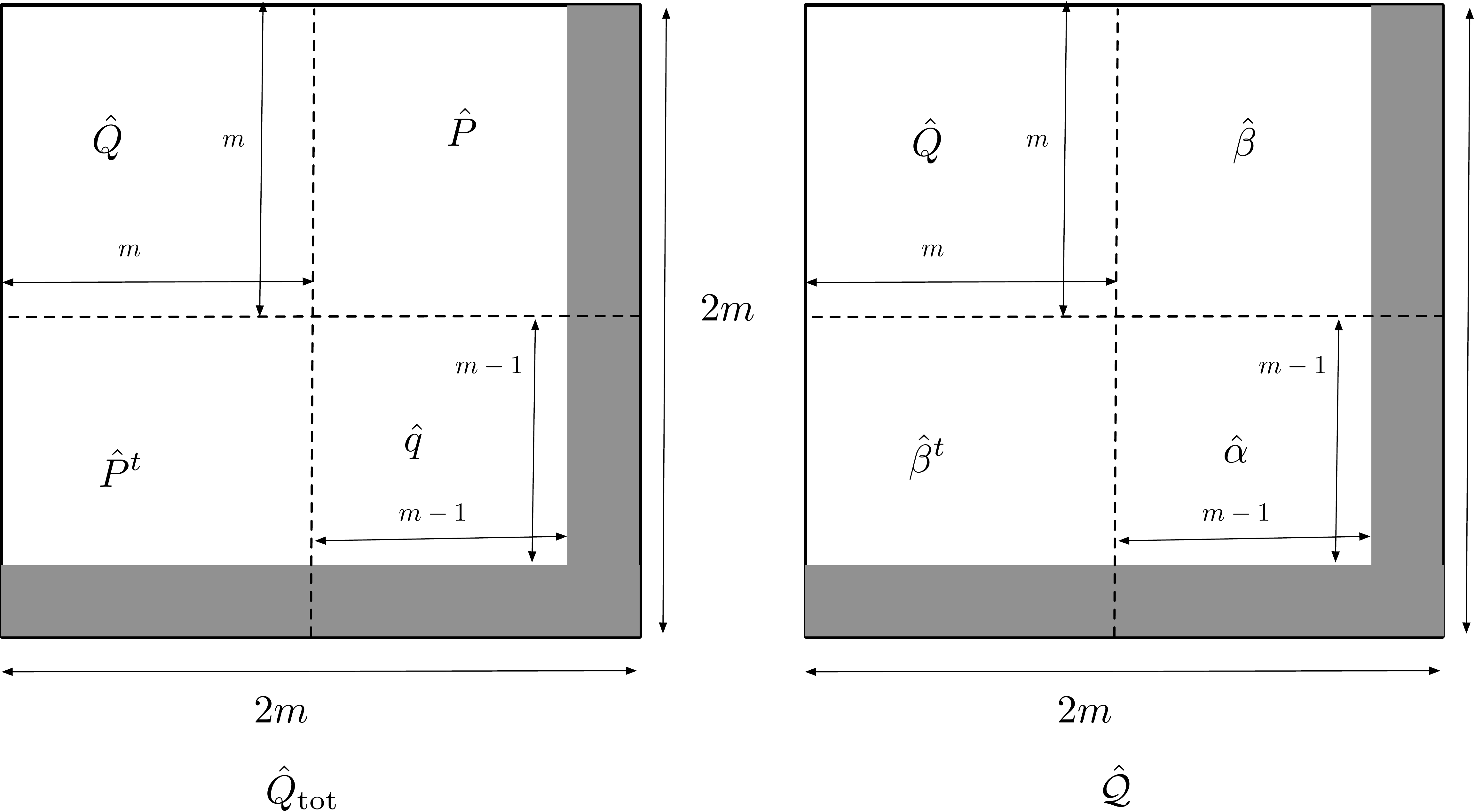}
  \caption{Parameterization of the extended Parisi's matrix $\hat{Q}_{\rm tot}$
    and its dimension-less version
    $\hat{\cal Q}$. The elements of the two matrices are
    related via \eq{eq-sm-def-alpha-beta}.}
	       \label{fig:parisi_extended_Qab}
\end{figure}

%Note that the 2nd equation which defines $Q_{ab}$ is the same as \eq{eq-sm-def-Qab}(using $M=d$).
For convenience we define a combined matrix as
shown in Fig. \ref{fig:parisi_extended_Qab} and call it as $\hat{Q}_{\rm tot}$
of size $2m \times 2m$.
Because of the sum rules \eq{eq-sm-sum-rule},
 $\hat{Q}_{\rm tot}$ can be parameterized completely
by a smaller matrix $\hat{Q}_{\rm tot}^{2m,2m}$ which is defined by
subtracting the $2m$-th row and column (shaded region in Fig.~\ref{fig:parisi_extended_Qab}).

We expect the $\rho(\ovr,\ovS)$ of the glass states
which keep the translational and orientational invariance of the liquid can be
parameterized solely by the glass order parameters as,
\beq
\rho(\ovr,\ovS)=\rho(\hat{Q}_{\rm tot})
\eeq
Now we wish to find the exact expression of the free-energy $-\beta F/N$ of the molecular liquid of the anisotropic particles in terms of the glass order parameters ${\cal Q}_{ab}$ combining the result of the simple spheres  \cite{charbonneau2014exact}
and a recent work on vectorial spin systems reported in Ref.  \cite{yoshino2018}.

The 1st step is to change the integration variables in
\eq{eq-sm-free-energy-functional-nreplica-anistropic-particle} from $\ovr$, $\ovS$ to $\hat{Q}_{\rm tot}$.
Because of the decomposition \eq{eq-sm-decomposition-particle-coordinate}
let us change integration variables as,
$\int_{\ovr} \ldots =\int d^{d}R\int {\cal D}\ovu \ldots$
with ${\cal D}\ovu \equiv d^{d}\ovu m^{d}\delta(\sum_{a=1}^{m}\vu^{a})$.
Subsequently let us change the integration variables from $(\ovu,\ovS)$ to
$\hat{Q}_{\rm tot}$ defined in \eq{eq-sm-def-qab} introducing the Jacobian
\beqn
j(\hat{Q}_{\rm tot})&\equiv& \int {\cal D}\ovu d\ovS \prod^{1,m}_{a \leq b} \delta\left(q_{ab}-\sum_{\mu=1}^{d}(u^{a})^{\mu}(u^{b})^{\mu}\right)
\prod^{1,m}_{a \leq b} \delta\left(Q_{ab}-\frac{1}{d}\sum_{\mu=1}^{d}(S^{a})^{\mu}(S^{b})^{\mu}\right)\nonumber\\
&& \times \prod_{a=1}^{m}\prod_{b=1}^{m} \delta\left(P_{ab}-\frac{1}{\sqrt{d}}\sum_{\mu=1}^{d}(u^{a})^{\mu}(S^{b})^{\mu}\right)
\eeqn

Now following essentially the same steps in (332)-(348) of  \cite{yoshino2018}
(see also  \cite{kurchan2012exact,kurchan2013exact,charbonneau2014exact}),
we find the entropic part of the free-energy functional
, which is the 1st term on the r.~h.~s. of 
\eq{eq-sm-free-energy-functional-nreplica-anistropic-particle}, can be expressed as
a functional of the glass order parameter ${\cal Q}$ as,
%
%down to \eq{eq-sm-free-energy-n-replicas}
%is much the same here so we do not repeat those.
\begin{eqnarray}
-\frac{\beta F_{\rm ent}}{N}=
  1-\ln \rho  + d \ln m
  +\frac{(2m-1)d}{2}\ln \left(\frac{2\pi e}{d}\right)
   +   \frac{d}{2}\ln {\rm det} \hat{Q}_{\rm tot}^{2m,2m}. \qquad
\end{eqnarray}
%The free-energy as a functional of the order parameters consists of two parts : the entropic term and the interaction term, which are the first two terms and last term on the right hand side of \eq{eq-sm-free-energy-n-replicas}.
%The entropic term is readily obtained as in  \cite{kurchan2012exact,kurchan2013exact,charbonneau2014exact} (see equation (1) of  \cite{charbonneau2014exact}),

Next let us analyze the interaction part of the free-energy, i.~e. the 2nd term
on the r.h.s of \eq{eq-sm-free-energy-functional-nreplica-anistropic-particle},
%\beq
%%V(r^{a}_{12},\hat{\vr}^{a}_{12}\cdot \vS^{a}_{1},\hat{\vr}^{a}_{12}\cdot \vS^{a}_{2},\vS^{a}_{1}\cdot\vS^{a}_{2})]\simeq
%%%V\left(|\vR_{12}+\frac{D}{\sqrt{d}}(\veta^{a}_{1}-\veta^{a}_{2})|,\hat{\vR}_{12}\cdot \vS^{a}_{1},\hat{\vR}_{12}\cdot \vS^{a}_{2},\vS^{a}_{1}\cdot\vS^{a}_{2}\right) 
%\eeq
%where we used
%\beq
%\hat{\vr}^{a}_{12}
%=\frac{\vR_{1}-\vR_{2}+(D/\sqrt{d})(\veta_{1}^{a}-\veta_{2}^{a})}{|\vR_{1}-\vR_{2}+(D/\sqrt{d})(\veta_{1}^{a}-\veta_{2}^{a})|}
%\underset{d \to \infty}{\to} \hat{\vR}_{12}=\frac{\vR_{1}-\vR_{2}}{|\vR_{1}-\vR_{2}|}
%\eeq
\beqn
-\frac{\beta F_{\rm int}}{N} &=&
%\frac{1}{2\rho}
%\int {\cal D}\oveta_{1}{\cal D}\oveta_{2}
%\int_{\vS_{1},\vS_{2}} \rho(\oveta_{1},\ovS_{1})\rho(\oveta_{2},\ovS_{2})
%\overline{f}(\oveta_{1}-\oveta_{2},\vS_{1},\vS_{2}) \nonumber \\
%&=&
\frac{1}{2\rho}
%\int {\cal D}\oveta_{1}{\cal D}\oveta_{2}
\int 
d\hat{Q}_{{\rm tot},1}d\hat{Q}_{{\rm tot},2}
j(\hat{Q}_{{\rm tot},1})j(\hat{Q}_{{\rm tot},2})
 \rho(\hat{Q}_{{\rm tot},1})
\rho(\hat{Q}_{{\rm tot},2})
\overline{f}(\hat{Q}_{{\rm tot},1},\hat{Q}_{{\rm tot},2}) \qquad
\eeqn
Assuming the orientationally and translationally invariant potential \eq{eq-sm-orientationally-invariant-potential} we can write,
\beqn
&&  \overline{f}(\hat{Q}_{{\rm tot},1},\hat{Q}_{{\rm tot},2}) \equiv
\int d^{d}R
\left \langle 
f_{m} \left( |\ovr_{1}-\ovr_{2}|^{2},
%  \left|\vR+\frac{D}{\sqrt{d}}(\veta^{a}_{1}-\veta^{a}_{2})\right|,
   \hat{\vR}\cdot \vS_{1}^{a},
   \hat{\vR}\cdot \vS_{2}^{a},
   \vS_{1}^{a}\cdot \vS_{2}^{a}\right) \right \rangle_{\hat{Q}_{\rm tot}}
 \\
&   = &\int_{0}^{\infty} dR R^{d-1} \int d\Omega_{d}
\left \langle    f_{m} (R^{2}+2R\hat{R}\cdot(\ovu_{1}-\ovu_{2})+(\ovu_{1}-\ovu_{2})^{2}, \hat{\vR}\cdot \vS_{1}^{a},
   \hat{\vR}\cdot \vS_{2}^{a},
   \vS_{1}^{a}\cdot \vS_{2}^{a}) \right \rangle_{\hat{Q}}   \nonumber \\
   &=& \int_{0}^{\infty} dR R^{d-1}
   \int \frac{d\ov{\lambda}}{2\pi}\frac{d\ov{\gamma}}{2\pi}\frac{d\ov{\gamma}'}{2\pi}\frac{d\ov{\kappa}}{2\pi}  \tilde{f}_{m}(\ov{\lambda},\ov{\gamma},\ov{\gamma}',\ov{\kappa})
\left \langle   e^{i\sum_{a=1}^{m}\kappa_{a}\vS_{1}^{a}\cdot\vS_{2}^{a}} 
   e^{i\sum_{a=1}^{m}\lambda_{a}(\vu_{1}^{a}-\vu_{2}^{a})^{2}}\right.
\nonumber \\
&&  \left.    \int d\Omega_{d}    e^{i\hat{R}\cdot \sum_{a=1}^{m}(2R\lambda_{a}(\vu_{1}^{a}-\vu_{2}^{a})+\gamma_{a}\vS_{1}^{a}+\gamma_{a}'\vS_{2}^{a}))} \right \rangle_{\hat{Q}_{\rm tot}}
   \eeqn
   Here $\int d^{d}R$ in an integration over the separation of the center of mass between the molecules
   $1$ and $2$ and $\hat{R}=\vR/R$ is a unit vector 
   whose direction represented by  the solid angle $\Omega$ is integrated over by $\int d\Omega \ldots$.
   Here we introduced
\begin{eqnarray}
&&  \langle \cdots \rangle_{\hat{Q}_{\rm tot}}
%  &\equiv&
    \equiv
    \int \prod_{l=1,2}\left \{
    d\ovu_{l} d\ovS_{l} \frac{1}{J(\hat{Q}_{{\rm tot},l})} \prod_{a \leq b}^{1,m}
\delta
\left((q_{l})_{ab}-\frac{1}{d}\sum_{\mu=1}^{d}((u_{l})^{a})^{\mu}((u_{l})^{b})^{\mu}\right) \prod^{1,m}_{a \leq b} \delta\left((Q_{l})_{ab}-\frac{1}{d}\sum_{\mu=1}^{d}((S_{l})^{a})^{\mu}((S_{l})^{b})^{\mu}\right) \right. \nonumber \\
&& \left. \times  \prod_{a=1}^{m}\prod_{b=1}^{m} \delta\left((P_{l})_{ab}-\frac{1}{\sqrt{d}}\sum_{\mu=1}^{d}((u_{l})^{a})^{\mu}((S_{l})^{b})^{\mu}\right)
 \right \}   \cdots
\label{eq-sm-av-with-q}
\end{eqnarray}
We also introduced a Fourier transform,
\beq
f_{m}(R^{2}+\ov{y},\ov{x},\ov{x}',\ov{h})=
\int \frac{d\ov{\lambda}}{2\pi}\frac{d\ov{\gamma}}{2\pi}\frac{d\ov{\gamma}'}{2\pi}\frac{d\ov{\kappa}}{2\pi}
e^{i\sum_{a=1}^{m}(\lambda_{a} y_{a}+\gamma_{a}x_{a}+\gamma'_{a}x'_{a}+\kappa_{a}h_{a})}
\tilde{f}_{m}(\ov{\lambda},\ov{\gamma},\ov{\gamma}',\ov{\kappa}).
\eeq

The integration over the solid angle can be evaluated
as follows. For arbitrary $d$-dimensional vectors
${\bf A}_{a}$ we have,
\beq
\int d\Omega_{d} e^{i\hat{R}\cdot \sum_{a=1}^{m}{\bf A}_{a}}
=\Omega_{d}\left \{
1-\frac{1}{2d}\sum_{\mu=1}^{d}\sum_{a,b=1}^{m} A^{\mu}_{a}A^{\mu}_{b}+O(1/d)
\right \}
\eeq
where $\Omega_{d}$ is the solid angle in the $d$-dimensional space.
This can be seen by noting $\langle \hat{R}^{\mu} \rangle_{\Omega}=0$,
$\langle \hat{R}^{\mu}\hat{R}^{\nu} \rangle_{\Omega}=(1/d)\delta_{\mu\nu}$,...
where we defined the average over the solid angle $\langle \ldots \rangle \equiv (1/\Omega_{d})\int d\Omega_{d} \ldots $. As in the case of
the $M$-component spin system in $M \to \infty$ limit, 
we assume that different components $\hat{R}^{\mu}$ become independent from each other in $d \to \infty$ limit.

Using $A_{a}=2R\lambda_{a}(\vu_{1}^{a}-\vu_{2}^{a})+\gamma_{a}\vS_{1}^{a}+\gamma_{a}'\vS_{2}^{a})$ in the above formula we find,
\beqn
&& \left\langle e^{i\sum_{a=1}^{m}\kappa_{a}\vS_{1}^{a}\cdot\vS_{2}^{a}} 
e^{i\sum_{a=1}^{m}\lambda_{a}(\vu_{1}^{a}-\vu_{2}^{a})^{2}} \int d\Omega_{d} e^{i\hat{R}\cdot \sum_{a=1}^{m}[2R\lambda_{a}(\vu_{1}^{a}-\vu_{2}^{a})+\gamma_{a}\vS_{1}^{a}+\gamma_{a}'\vS_{2}^{a}]} \right\rangle_{\hat{Q}_{\rm tot}} \nonumber \\
&=& \exp \left(
i\sum_{a=1}^{m}\lambda_{a}((q_{1})_{aa}+(q_{2})_{aa})
-\frac{1}{2d}\sum_{a,b=1}^{m} (2R\lambda_{a})(2R\lambda_{b})((q_{1})_{ab}+(q_{2})_{ab}))
\right. \nonumber \\
&& \left.
-\frac{1}{d}\sum_{ab}(2R)\lambda_{a}(\gamma_{b}(P_{1})_{ab}-\gamma'_{b}(P_{2})_{ab})
-\frac{1}{2}\sum_{ab}\gamma_{a}\gamma_{b}(Q_{1})_{ab}-\frac{1}{2}\sum_{a,b=1}^{m}\gamma'_{a}\gamma'_{b}(Q_{2})_{ab}
-\frac{1}{2}\sum_{a,b=1}^{m}\kappa_{a}\kappa_{b}(Q_{1})_{ab}(Q_{2})_{ab}
\right) \qquad
\eeqn
and we obtain,
\beqn
&&  \ov{f}(\hat{Q}_{{\rm tot},1},\hat{Q}_{{\rm tot},2})=
\Omega_{d}\int_{0}^{\infty} dRR^{d-1} \int d\ov{y} d\ov{x} d\ov{x'} d\ov{h}
\left \{
\exp \left[
-\sum_{a=1}^{m}((q_{1})_{aa}+(q_{2})_{aa})\frac{\partial}{\partial y_{a}}
+\frac{1}{2d}\sum_{a,b=1}^{m} \frac{\partial^{2}}{\partial y_{a}\partial y_{b}}(2R)^{2}((q_{1})_{ab}+(q_{2})_{ab}))
\right. \right. \nonumber \\
&& \left. \left.
+\frac{2R}{\sqrt{d}}\sum_{ab}\frac{\partial^{2}}{\partial y_{a}\partial x_{b}}(P_{1})_{ab}
-\frac{2R}{\sqrt{d}}\sum_{ab}\frac{\partial^{2}}{\partial y_{a}\partial x'_{b}}(P_{2})_{ab}+\frac{1}{2}\sum_{a,b=1}^{m}\frac{\partial^{2}}{\partial x_{a}\partial x_{b}}(Q_{1})_{ab}+\frac{1}{2}\sum_{a,b=1}^{m}\frac{\partial^{2}}{\partial x'_{a}\partial x'_{b}}(Q_{2})_{ab}
\right. \right.\\  \nonumber
 && \left.  \left. 
+\frac{1}{2}\sum_{a,b=1}^{m}(Q_{1})_{ab}(Q_{2})_{ab}\frac{\partial^{2}}{\partial h_{a}\partial h_{b}} \right]
\prod_{a=1}^{m} \left \{
\delta(y_{a})\delta(x_{a})\delta(x'_{a})\delta(h_{a})\right\}
\right \} 
 f_{m}(R^{2}+\ov{y},\ov{x},\ov{x}',\ov{h})  \nonumber \\
&& =\Omega_{d}\int_{0}^{\infty} dRR^{d-1}
\exp \left[
  \sum_{a=1}^{m}((q_{1})_{aa}+(q_{2})_{aa})\frac{\partial}{\partial y_{a}}
  +\frac{1}{2d}\sum_{a,b=1}^{m} \frac{\partial^{2}}{\partial y_{a}\partial y_{b}}(2R)^{2}((q_{1})_{ab}+(q_{2})_{ab}))
  \right. \nonumber \\
&&  \left. \left. +\frac{2R}{\sqrt{d}}\sum_{ab}\frac{\partial^{2}}{\partial y_{a}\partial x_{b}}(P_{1})_{ab}
  -\frac{2R}{\sqrt{d}}\sum_{ab}\frac{\partial^{2}}{\partial y_{a}\partial x'_{b}}(P_{2})_{ab}+\frac{1}{2}\sum_{a,b=1}^{m}\frac{\partial^{2}}{\partial x_{a}\partial x_{b}}(Q_{1})_{ab}+\frac{1}{2}\sum_{a,b=1}^{m}\frac{\partial^{2}}{\partial x'_{a}\partial x'_{b}}(Q_{2})_{ab}
 \right. \right. \nonumber \\
 && \left.  \left. 
+\frac{1}{2}\sum_{a,b=1}^{m}(Q_{1})_{ab}(Q_{2})_{ab}\frac{\partial^{2}}{\partial h_{a}\partial h_{b}}\right]f_{m}(R^{2}+\ov{y},\ov{x},\ov{x}',\ov{h}) \right |_{\ov{y}=\ov{x}=\ov{x}'=\ov{h}=0} 
\eeqn

Now we have to evaluate the integration $\int dR R^{d-1} \ldots$. In order to take
$d\to \infty$ limit, it is useful to change the integration variable from $R$
to a scaled variable $\xi$  \cite{charbonneau2014exact},
\beq
R\equiv D\left(1+\frac{\xi}{d}\right)
\eeq
with which we can write
\beq
\int_{0}^{\infty} dR R^{d-1} \underset{d \to \infty}{\rightarrow} \frac{D^{d}}{d}\int_{-\infty}^{\infty} d\xi e^{\xi}
\eeq
then we obtain
\beqn
&&  \overline{f}(\hat{Q}_{{\rm tot},1},\hat{Q}_{{\rm tot},2}) \underset{d\to\infty}{=}
\frac{D^{d}}{d}\Omega_{d}\int_{-\infty}^{\infty} d\xi e^{\xi}
\exp \left[
  %-\frac{1}{4} ((\Delta_{1})_{ab}+(\Delta_{2})_{ab}))  \frac{\partial^{2}}{\partial \xi_{a}\partial \xi_{b}}
  \frac{1}{2}\sum_{a}((\alpha_{1})_{aa}+(\alpha_{2})_{aa}))  \frac{\partial}{\partial \xi_{a}}
+ \frac{1}{2} \sum_{ab}((\alpha_{1})_{ab}+(\alpha_{2})_{ab}))  \frac{\partial^{2}}{\partial \xi_{a}\partial \xi_{b}}
  \right. \nonumber \\
&&  \left. +\sum_{ab}\left ( (\beta_{1})_{ab}\frac{\partial^{2}}{\partial \xi_{a}\partial x_{b}}-(\beta_{2})_{ab}\frac{\partial^{2}}{\partial \xi_{a}\partial x' _{b}} \right)
\right. \nonumber \\
 && \left.  \left. 
+\frac{1}{2}\sum_{ab}(Q_{1})_{ab}\frac{\partial^{2}}{\partial x_{a}\partial x_{b}}+\frac{1}{2}\sum_{ab}(Q_{2})_{ab}\frac{\partial^{2}}{\partial x_{a}\partial x_{b}}+\frac{1}{2}\sum_{ab}(Q_{1})_{ab}(Q_{2})_{ab}\frac{\partial^{2}}{\partial h_{a}\partial h_{b}}\right]f_{m}\left(D^{2}\left(1+\frac{\ov{\xi}}{d}\right)^{2},\ov{x},\ov{x}',\ov{h}\right) \right |_{\ov{\xi}=\xi,\ov{y}=\ov{x}=\ov{x}'=\ov{h}=0} \nonumber \\
&& =\frac{D^{d}}{d}\Omega_{d}\int_{-\infty}^{\infty} d\xi e^{\xi}
\prod_{a,b=1}^{m}
\exp \left[
  -\frac{1}{4} ((\Delta_{1})_{ab}+(\Delta_{2})_{ab}))  \frac{\partial^{2}}{\partial \xi_{a}\partial \xi_{b}}
%  ((\alpha_{1})_{aa}+(\alpha_{2})_{aa}))  \frac{\partial^{2}}{\partial \xi_{a}\partial \xi_{b}}
%+ \frac{1}{2} ((\alpha_{1})_{ab}+(\alpha_{2})_{ab}))  \frac{\partial^{2}}{\partial \xi_{a}\partial \xi_{b}}
%  \right. \nonumber \\
%  &&  \left.
%  -\sum_{ab}\left ( (\beta_{1})_{ab}\frac{\partial}{\partial x_{b}}-(\beta_{2})_{ab}\frac{\partial}{\partial x' _{b}} \right)
\right. \nonumber \\
 && \left.  \left. 
+\frac{1}{2}(Q_{1})_{ab}\frac{\partial^{2}}{\partial x_{a}\partial x_{b}}+\frac{1}{2}(Q_{2})_{ab}\frac{\partial^{2}}{\partial x_{a}\partial x_{b}}+\frac{1}{2}(Q_{1})_{ab}(Q_{2})_{ab}\frac{\partial^{2}}{\partial h_{a}\partial h_{b}}\right]f_{m}\left(D^{2}\left(1+\frac{\ov{\xi}}{d}\right)^{2},\ov{x},\ov{x}',\ov{h}\right) \right |_{\ov{\xi}=\xi,\ov{y}=\ov{x}=\ov{x}'=\ov{h}=0}
\label{eq-sm-fint-nreplica-anistropic-particle}
\eeqn
where we used $R \to D$ in $d\to \infty$ for finite $\xi$ and introduced scaled order parameters,
\beq
\alpha_{ab} \equiv \frac{d}{D^{2}}q_{ab} \qquad \beta_{ab}=\frac{\sqrt{d}}{D}P_{ab}
\label{eq-sm-def-alpha-beta}
\eeq
and
\beq
\Delta_{ab}\equiv \alpha_{aa}+\alpha_{bb}-2\alpha_{ab} \qquad 
\eeq
Correspondingly the sum rule \eq{eq-sm-sum-rule} becomes,
\beq
\sum_{b=1}^{m}\alpha_{ab}=\sum_{a=1}^{m}\alpha_{ab}=0 \qquad
\sum_{b=1}^{m}\beta_{ab}=\sum_{a=1}^{m}\beta_{ab}=0
\label{eq-sm-sum-rule-2}
\eeq

Let us explain the derivation of the last equation of \eq{eq-sm-fint-nreplica-anistropic-particle}.
The integral over $\xi$ can be rewritten as follows,
\beqn
&& \int_{-\infty}^{\infty} d\xi e^{\xi} \left.
e^{\sum_{a}\alpha_{aa} \frac{\partial}{\partial \xi_{a}}
  +  \sum_{ab}\alpha_{ab}  \frac{\partial^{2}}{\partial \xi_{a}\partial \xi_{b}}
  + \sum_{ab}\beta_{ab} \frac{\partial^{2}}{\partial \xi_{a}\partial x_{b}}
} A(\ov{\xi})\right |_{\ov{\xi}=0}  \nonumber \\
&=&  \int_{-\infty}^{\infty} d\xi e^{\xi}
\left. 
e^{\alpha_{\rm d}
  (\sum_{a}\frac{\partial}{\partial \xi_{a}}+1)\sum_{b}\frac{\partial}{\partial \xi_{b}}}
e^{- \frac{1}{2} \sum_{ab}\Delta_{ab}  \frac{\partial^{2}}{\partial \xi_{a}\partial \xi_{b}}}
e^{-\sum_{a}\beta_{ab} \sum_{b}\frac{\partial}{\partial x_{b}}}
A(\ov{\xi})\right |_{\ov{\xi}=0} \nonumber \\
&=&
\int_{-\infty}^{\infty} d\xi e^{\xi}
\left. 
e^{- \frac{1}{2} \sum_{ab}\Delta_{ab}  \frac{\partial^{2}}{\partial \xi_{a}\partial \xi_{b}}}
A(\ov{\xi})\right |_{\ov{\xi}=0}
\eeqn
where $A(\ov{\xi})$ represents the operand of the
differential operators $\partial_{\xi_{a}}$.
In the 2nd equation we assumed that the diagonal elements of the $\hat{\alpha}$ matrix is a constant, say $\alpha_{aa}=\alpha_{\rm d}$  \cite{charbonneau2014exact}, which is independent of replicas $a$. This holds for the usual Parisi's replica symmetry breaking ansatz  \cite{parisi1979infinite}. 
In the 2nd equation we repeatedly performed integrations by parts over $\xi$ concerning
the factor  which depends on $\beta_{ab}$.
Also in the last step we repeated integrations by parts over $\xi$ and
used $\int_{-\infty}^{\infty} d\xi e^{\xi} (\frac{d}{d\xi}+1) A(\xi)=0$.
In the last step we also used the sum rule \eq{eq-sm-sum-rule-2} by which we find the term with $\beta_{ab}$ vanishes.

Assuming that $\hat{Q}_{{\rm tot},1}^{*}=\hat{Q}_{{\rm tot},2}^{*}$
at the saddle point,
which is defined by normalization conditions
$\rho=\int d\hat{Q}_{\rm tot}\rho(\hat{Q}_{\rm tot})$
 \cite{kurchan2013exact,yoshino2018}, 
we find the interaction part of the free-energy as,
\beqn
-\frac{\beta F_{\rm int}}{N} &=&
\frac{d}{2}\hat{\varphi}
\int_{-\infty}^{\infty}d\xi e^{\xi} e^{\frac{1}{2}\sum_{a,b=1}^{m} {\cal D}_{ab}}
\left[
  \left.  \prod_{a=1}^{m}e^{-\beta V (D(1+\xi_{a}/d),x_{a},x'_{a},y_{a})} \right |_{\substack{ \{\xi_{a}=\xi\} \\ \{x_{a},x'_{a},h_{a}=0\} }}-1
  \right]
\eeqn
where we introduced
\beq
    {\cal D}_{ab}=-\Delta_{ab}\frac{\partial^{2}}{\partial \xi_{a}\partial \xi_{b}}
%    +\beta_{ab}
%    \left (
%    \frac{\partial}{\partial \xi_{a}}+\frac{\partial}{\partial \xi_{b}}
%    \right)
%        \left (
%    \frac{\partial}{\partial x_{b}}-\frac{\partial}{\partial x' _{b}}
%    \right)
+Q_{ab}\left (
\frac{\partial^{2}}{\partial x_{a}\partial x_{b}}
+\frac{\partial^{2}}{\partial x'_{a}\partial x'_{b}}
\right)
+Q_{ab}^{2}\frac{\partial^{2}}{\partial h_{a}\partial h_{b}}
\label{eq-sm-D}
\eeq
and
\beq
\hat{\varphi}\equiv \rho 2^{d}\varphi/d \qquad \varphi=\rho \frac{\Omega_{d}}{d}(D/2)^{d}
\eeq

Including  the entropic part of the free-energy we finally obtain,
\begin{eqnarray}
  -\beta m \phi_{m}[\hat{\cal Q}] &=& c_{\rm nt}
   +   \frac{d}{2}\ln {\rm det} \hat{\cal Q}^{2m,2m}
   -\frac{d}{2}\hat\varphi {\cal F}_{\rm int}[\hat{\cal Q}]
   \label{eq-sm-replica-free-energy-final}
\end{eqnarray}
where
\beq
c_{\rm nt}= 1-\ln \rho  + d \ln m
    +\frac{(m-1)d}{2}\ln \left(\frac{2\pi eD^{2}}{d^{2}}\right)
    +\frac{d}{2}m\ln \left( \frac{2\pi e}{d}\right)
    \label{eq-sm-cnt}
\eeq
and
\begin{eqnarray}
&&   -{\cal F}_{\rm int}[\hat{\cal Q}] \equiv  
  \int_{-\infty}^{\infty} d\xi e^{\xi}  e^{
    \frac{1}{2}\sum_{a,b=1}^{m} {\cal D}_{ab}
        } 
      \left[
       \prod_{a=1}^{m}  e^{-\beta V(D(1+\frac{\xi_{a}}{d}),x_{a},x'_{a},h_{a})}
     |_{\substack{ \{\xi_{a}=\xi\} \\ \{x_{a},x'_{a},h_{a}=0\} }}  -1 \right]
     \label{eq-sm-f-int-anisotropic-particle}
\end{eqnarray}
Here we introduced a matrix $\hat{\cal Q}$ which
combines the matrices $Q_{ab}$, $\alpha_{ab}$, $\beta_{ab}$
as shown in Fig. \ref{fig:parisi_extended_Qab}.

\section{1RSB ansatz}

\begin{figure}[t]
  \includegraphics[width=0.4\textwidth]{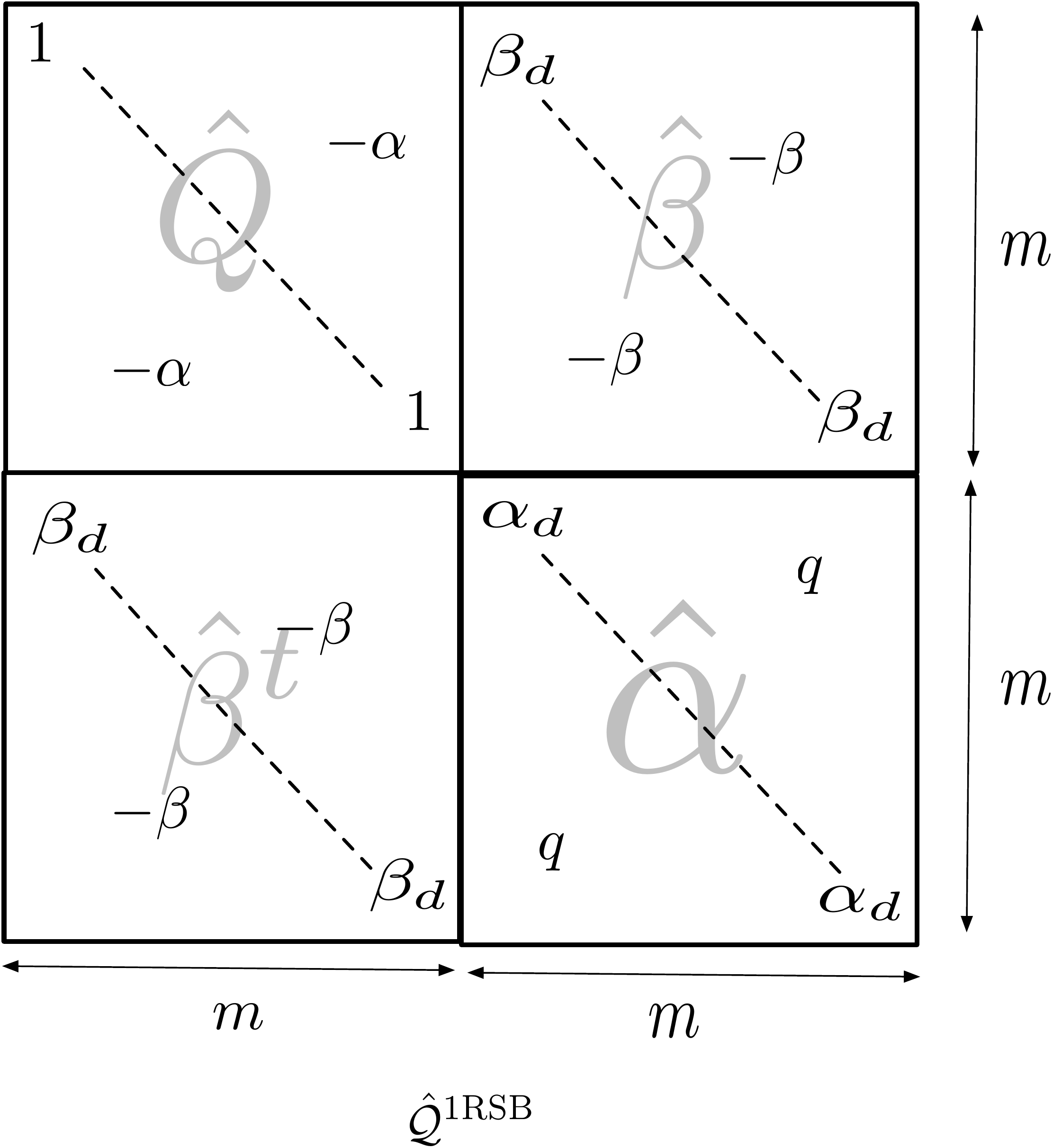}
  \caption{Parameterization of the 1RSB matrix}
	       \label{fig:parisi_extended_Qab_1RSB}
\end{figure}

Let us consider the 1RSB ansatz which amount to assume
the matrix $\hat{Q}$ in the form $\hat{Q}^{\rm 1RSB}$ shown in
Fig.~\ref{fig:parisi_extended_Qab_1RSB}.
The $m\times m$ sub-matrices $\hat{\alpha}$, $\hat{\beta}$ and $\hat{Q}$ can be
expressed as,
\beq
\hat{\alpha}=(m I-1)\alpha \qquad
\hat{\beta}=(mI-1)\beta \qquad
\hat{Q}=(1-q)I+q
\eeq
where $I$ is the identity matrix of size $m\times m$.
The sum rules \eq{eq-sm-sum-rule-2} which reflect the translational invariance
imply the diagonal elements
of the sub-matrices $\hat{\alpha}$ and $\hat{\beta}$ can be expressed as,
\beq
\alpha_{d}=(m-1)\alpha \qquad \beta_{d}=(m-1)\beta.
\eeq
We also note that the matrix
$\hat{\Delta}$ defined as $\Delta_{ab}=\alpha_{aa}+\alpha_{bb}-2\alpha_{ab}$
can be written as,
\beq
\hat{\Delta}=\Delta(1-I)
\eeq
with
\beq
\Delta=2(\alpha_{\rm d}+\alpha)=2m\alpha
\label{eq-sm-Delta-alpha}
\eeq

\subsection{Some useful formulae}

The sub-matrices $\hat{\alpha}$, $\hat{\beta}$ and $\hat{Q}$
are all $m\times m$ matrix of the following form in the 1RSB ansatz,
        \beq
        \hat{A}=a_{1}I + a_{2}.
        \eeq
Let us display below some useful formulae,
        \beq
        \hat{A}^{-1}=\frac{1}{a_{1}}I-\frac{a_{2}}{a_{1}(a_{1}+ma_{2})}
\qquad
    {\rm det} A=a_{1}^{m-1}(a_{1}+m a_{2})
    \label{eq-sm-1RSB-algebra}
    \eeq

    Note however that $\hat{\alpha}$ and $\hat{\beta}$ are not invertible
    because they respect the sum rule
    \eq{eq-sm-sum-rule-2} which implies $a_{1}+m a_{2}=0$.
    In the following we use a
    trick to circumvent this is  to make a shift $a_{1} \to a_{1}+\epsilon$ and then
    take $\epsilon \to 0$ in the end of computations (see supplementary
    information of  \cite{RUYZ14} sec. I B).

\subsection{Entropic part of the free-energy}

First let us analyze the entropic contribution to the
free-energy \eq{eq-sm-replica-free-energy-final} within the 1RSB ansatz.
Using the cofactor expansion for inverse matrices we can write
\beq
    {\rm det} (\hat{\cal Q})^{2m,2m}=
    {\rm det} \hat{\cal Q}
      (\hat{\cal Q})^{-1}_{2m,2m}
    \eeq
    Given the structure of the matrix (See Fig.~\ref{fig:parisi_extended_Qab_1RSB}) we can write (see  \cite{james2007gentle} Eq. (3.147)),
    \beq
        {\rm det} \hat{\cal Q}=
        {\rm det} \hat{Q} \;{\rm det} Z
        \eeq
        and (see  \cite{james2007gentle} Eq. (3.145))
        \beq
     (\hat{\cal Q})^{-1}_{2m,2m}=\hat{Z}^{-1}_{m,m}
        \eeq
        Here we introduced,
        \beq
        \hat{Z}=\hat{\alpha}-\hat{\beta}^{\rm t}\hat{Q}^{-1}\hat{\beta}
        \label{eq-sm-matrix-z}
        \eeq
        Now by writing
        \beq
        \hat{Z}=z_{1}I+z_{2}
        \label{eq-sm-matrix-z-2}
        \eeq
         and using the formulae \eq{eq-sm-1RSB-algebra}
        we find
                \beq
            {\rm det} (\hat{\cal Q})^{2m,2m}
            = {\rm det} (\hat{Q}Z^{-1}_{m,m}
            =(1-q)^{m-1}[1+(m-1)q]z_{1}^{m-1}
            \left( 1+(m-1)\frac{z_{2}}{z_{1}} \right)
        \eeq
On the other hand, using \eq{eq-sm-matrix-z}, \eq{eq-sm-matrix-z-2} and \eq{eq-sm-1RSB-algebra} we find after some algebra
\beq
z_{1}=m\left(\alpha-\frac{m\beta^2}{1-q}\right)
\qquad z_{2}=-\left(\alpha-\frac{m\beta^2}{1-q}\right)
        \eeq
        Collecting the above results, we find the
        following factor which is
        essential in the
        entropic contribution to the free-energy \eq{eq-sm-replica-free-energy-final},
\beqn
\ln {\rm det} (\hat{\cal Q}^{\rm 1RSB})^{2m,2m}=
(m-1)\ln\left(\frac{\Delta}{2}-\frac{m^{2}\beta^{2}}{1-q}\right)-\ln m\nonumber \\
+\ln (1+(m-1)q)+(m-1)\ln(1-q) \qquad
\label{eq-sm-detQ-1RSB}
\eeqn
where we used $\Delta=2m\alpha$ given in \eq{eq-sm-Delta-alpha}.

\subsection{Interaction part of the free-energy}

Within the 1RSB ansatz the differential operator ${\cal D}$
in \eq{eq-sm-D} becomes,
\beq
    {\cal D}_{ab}=D_{1}\delta_{ab}+D_{2}.
    \label{eq-sm-D-1RSB}
\eeq
with
\beqn
&& D_{1}=\Delta \partial_{\xi}^{2}
+(1-q)(\partial_{x}^{2}+\partial_{x'}^{2})+(1-q^{2})\partial_{h}^{2} \qquad \\
&& D_{2}=-\Delta \partial_{\xi}^{2}
+q(\partial_{x}^{2}+\partial_{x'}^{2})
+q^{2}\partial_{h}^{2} \qquad
    \label{eq-sm-D1-D2}
\eeqn
Then the functional ${\cal F}_{\rm int}[\hat{\cal Q}]$
\eq{eq-sm-f-int-anisotropic-particle}
for the interaction part of the free-energy becomes,
\beqn
-{\cal F}_{\rm int}[\hat{\cal Q}^{\rm 1RSB}]
&=& \int_{-\infty}^{\infty}d\xi e^{\xi}
e^{\frac{1}{2}{\cal D}_{2}}
\left. \left( g^{m}(\xi,x,x',h) -1 \right) \right |_{x=x'=h=0}
    \label{eq-sm-Fint-1RSB-0}
\eeqn
where we introduced
\beqn
g(\xi,x,x',h)=e^{\frac{1}{2}{\cal D}_{1}} e^{-\beta V(\xi,x,x',h)}
\label{eq-sm-def-g}
\eeqn

The functional ${\cal F}_{\rm int}[\hat{\cal Q}^{\rm 1RSB}]$ can be cast into more convenient form.
To this end we use the formula  \cite{duplantier1981comment}
\beq
\exp\left(\frac{a}{2}\frac{\partial^{2}}{\partial h^{2}}\right)A(h)=\gamma_{a} \otimes A(h)
\label{eq-sm-formula1}
\eeq
where $\gamma_{a}(x)$ is a Gaussian with zero mean and variance $a$,
\beq
\gamma_{a}(x)=\frac{1}{\sqrt{2\pi a}}e^{-\frac{x^{2}}{2a}},
\eeq
by which we write a convolution of a function $A(x)$ with the Gaussian as,
\beq
\gamma_{a}\otimes A(x) \equiv \int dy \frac{e^{-\frac{y^{2}}{2a}}}{\sqrt{2\pi a}}A(x-y)=\int {\cal D}zA(x-\sqrt{a}z)
\label{eq-sm-formula2}
\eeq
where
\beq
\int \cD z \ldots \equiv \int
dz \frac{e^{-\frac{z^{2}}{2}}}{\sqrt{2\pi}} \cdots
\eeq
Using the above formula we find,
\beqn
-{\cal F}_{\rm int}[\hat{\cal Q}^{\rm 1RSB}]
&=& \int_{-\infty}^{\infty}d\xi e^{\xi}
e^{-\frac{\Delta}{2}\partial_{\xi}^{2}}
\int {\cal D}z_{x}{\cal D}z_{x'}{\cal D}z_{h}
\left. \left( g^{m}(\xi,x-\sqrt{q}z_{x},x'-\sqrt{q}z_{x'},h-\sqrt{q^{2}}z_{h}) -1 \right) \right |_{x=x'=h=0}  \nonumber \\
&=& \int_{-\infty}^{\infty}d\xi e^{\xi-\Delta/2}
\left. \left(
\int {\cal D}z_{x}{\cal D}z_{x'}{\cal D}z_{h}
g^{m}(\Xi)
\right |_{\Xi=(\xi,\sqrt{q}z_{x},\sqrt{q}z_{x'},\sqrt{q^{2}}z_{h})}
-1 \right) 
\label{eq-sm-Fint-1RSB}
\eeqn
In the last equation we repeatedly performed integrations by parts over $\xi$.
Similarly we find \eq{eq-sm-def-g} becomes,
\beqn
g(\xi,x,x',h)&=&e^{\frac{1}{2}{\cal D}_{1}}  e^{-\beta V(\xi,x,x',h)} \nonumber \\
&=& \int {\cal D}z_{\xi} {\cal D}z_{x}{\cal D}z_{x'}{\cal D}z_{h}e^{-\beta V(\xi-\sqrt{\Delta}z_{\xi},x-\sqrt{1-q}z_{x},x'-\sqrt{1-q}z_{x},h-\sqrt{1-q^{2}}z_{h})}
\label{eq-sm-g-integral-form}
\eeqn

\subsection{Replicated free-energy and Franz-Parisi potential}

To sum up we obtain the free-energy within the 1RSB ansatz as,
\begin{eqnarray}
  -\beta m  \phi(\Delta,q,\beta) &=& c_{\rm nt}
  +   \frac{d}{2}\left[
    (m-1)\ln\left(\frac{\Delta}{2}-\frac{m^{2}\beta^{2}}{1-q}\right)-\ln m 
%(1-q)^{m-1}[1+(m-1)q]z_{1}^{m-1}
    %            \left( 1+(m-1)\frac{z_{2}}{z_{1}} \right)
    \right]
%  \ln {\rm det} (\hat{\cal Q}^{\rm 1RSB})^{2m,2m} 
+\frac{d}{2} \hat\varphi {\cal F}_{\rm int}(\Delta,q)
  %\hat\varphi   \int_{-\infty}^{\infty}d\xi e^{\xi}
%e^{\frac{1}{2}{\cal D}_{2}}
%\left. \left( g^{m}(\xi,x,x',h) -1 \right) \right |_{x=x'=h=0}
   \label{eq-sm-replica-free-energy-1RSB}
\end{eqnarray}
with
\beq
    {\cal F}_{\rm int}(\Delta,q)=
     \int_{-\infty}^{\infty}d\xi e^{\xi}
 \left( \int {\cal D}z_{x}{\cal D}z_{x'}{\cal D}z_{h}
 \left.g^{m}(\Xi) \right |_{\Xi=(\xi+\Delta/2,\sqrt{q}z_{x},\sqrt{q}z_{x'},\sqrt{q^{2}}z_{h})}
 -1 \right)
 \label{eq-sm-Fint-1RSB-v2}
\eeq
and $c_{\rm nt}$ given in \eq{eq-sm-cnt} and $g(\xi,x,x',h)$ given in \eq{eq-sm-g-integral-form}.

Close to $m=1$ it is useful to expand the free-energy in power series
of $s=1-m$ which yields,
\beq
-\beta (1+s)\phi_{1+s}(\Delta,q,\beta)=
-\beta \phi_{1}-\beta \nu_{\rm FP}(\Delta,q,\beta)s+O(s^{2})
\label{eq-sm-def-FP-0}
\eeq
with the Franz-Parisi potential,
\beqn
-\frac{2}{d}\beta \nu_{\rm FP} (\Delta,q,\beta)
&=&1+\ln \left[ \left(\frac{2\pi e}{d}\right)^{2}\frac{D^{2}}{d}\right]
+\ln (1-q)+q + \ln \left( \frac{\Delta}{2}-\frac{\beta^{2}}{1-q}\right) \nonumber \\
&& +\hat\varphi   \int_{-\infty}^{\infty}d\xi e^{\xi-\Delta/2}
\int {\cal D}z_{x}{\cal D}z_{x'}{\cal D}z_{h}
\left.  g(\Xi)
\ln g(\Xi)\right |_{\Xi=(\xi,\sqrt{q}z_{x},\sqrt{q}z_{x'},\sqrt{q^{2}}z_{h})}
\label{eq-sm-FP-1RSB}
\eeqn
We note that $\phi_{1}=\lim_{m \to 1}\phi_{m}(\Delta,q,\beta)$ is independent of
the order parameters $(\Delta,q,\beta)$ as can be checked using
\eq{eq-sm-detQ-1RSB} and     \eq{eq-sm-D-1RSB}-\eq{eq-sm-def-g}.

\subsection{Saddle point equations}

The saddle point equations for $\Delta$,$q$,$\beta$  are give by
\beq
0= \left. \frac{\partial \phi_{m}(\Delta,q,\beta)}{\partial \Delta}
\right |_{\Delta=\Delta^{*},q=q^{*},\beta=\beta^{*}}
\qquad
0=\left. \frac{\partial \phi_{m}(\Delta,q,\beta)}{\partial q}
\right |_{\Delta=\Delta^{*},q=q^{*},\beta=\beta^{*}}
\qquad
0=\left. \frac{\partial \phi_{m}(\Delta,q,\beta)}{\partial \beta}
\right |_{\Delta=\Delta^{*},q=q^{*},\beta=\beta^{*}}
|_{\Delta=\Delta^{*},q=q^{*}}
\eeq
We immediately find
\beq
\beta^{*}=0
\eeq
at the saddle point. Then the remaining saddle point equations
are found as,
\beqn
&& \frac{1}{\varphi}=\left. \frac{1}{1-m}
\left( \Delta\frac{\partial}{\partial \Delta} \right)
     {\cal G}_{m}(\Delta,q)
\right|_{\Delta=\Delta^{*},q=q^{*}}
     \nonumber \\
     && \frac{1}{\varphi}=
     \left.
     \frac{1}{1-m}\frac{1+(m-1)q}{mq}
     \left ((1-q) \frac{\partial}{\partial  (1-q)} \right){\cal G}_{m}(\Delta,q)
\right|_{\Delta=\Delta^{*},q=q^{*}}
     \label{eq-sm-SP-1RSB}
\eeqn
where we introduced
\beq
    {\cal G}_{m}(\Delta,q)=
  \int_{-\infty}^{\infty}d\xi e^{\xi}
 \left( \int {\cal D}z_{x}{\cal D}z_{x'}{\cal D}z_{h}
 \left.g^{m}(\Xi) \right |_{\Xi=(\xi+\Delta/2,\sqrt{q}z_{x},\sqrt{q}z_{x'},\sqrt{q^{2}}z_{h})}
 -\theta(\zeta) \right)
 \label{eq-sm-def-cal-G}
 \eeq
 by which we can rewrite ${\cal F}_{\rm int}$ in
 \eq{eq-sm-Fint-1RSB-v2} as
 \beq
 {\cal F}_{\rm int}(\Delta,q)=    {\cal G}_{m}(\Delta,q)-1
 \eeq

 Suppose that ${\cal G}_{m}(\Delta,q)$
scaled as
  \beq
      {\cal G}_{m}(\Delta,q)-{\cal G}_{m}(0,1)
      \propto C_{\Delta}\Delta^{\alpha_{\Delta}}
+C_{q}(1-q)^{\alpha_{q}}
\eeq
for small $\Delta$ and $1-q$. For example in the
case of the simple hardspheres $\alpha_{\Delta}=1/2$  \cite{parisi2010mean}.
Then we find the
saddle point values scales as,
\beq
\Delta^{*} \propto {\hat\varphi}^{-1/\alpha_{\Delta}}
\qquad
1-q^{*} \propto {\hat\varphi}^{-1/\alpha_{q}}
\label{eq-sm-SP-solution-large-phi}
\eeq
at large densities $\hat\varphi \gg 1$. Which in tern implies
\beq
    {\cal G}_{m}(\Delta^{*},q^{*})=
    {\cal G}_{m}(0,1)+O(\varphi^{-1})
    \eeq
    at large densities.

    \subsection{Outline of the analysis based on the 1RSB ansatz}

Let us outline the subsequent analysis.
After fixing the remaining order parameters $q$ and $\Delta$ at a non-trivial
saddle pointwith $\Delta < \infty$ and/or $q >0$,
we are left with the replica free-energy $\phi_{m}=\phi_{m}(\hat{Q^{*}}(m))$ which is still a function of the parameter $m$.
The thermodynamic free-energy (per particle) of the system is given by
$\lim_{m \to 1_{-}} \phi_{m}$.
Following the standard prescription \cite{Mo95}, the complexity or the configurational entropy
$\Sigma(f)$, by which the density of glassy states with free-energy (per particle)
in the range between $f$ and $f+df$ is given as $e^{N \sigma(f)}df$,
can be obtained through,
\beqn
&& \Sigma (m)=-\beta m  \phi_{m}+\beta m f (m) \nonumber \\
&& -\beta m f(m)= m \partial_{m} (-\beta m \phi_{m})
\eeqn
The thermodynamic free-energy (per particle) of the system is given by
\beq
\lim_{m \to 1} \phi_{m}=f(m^{*})/m^{*}-k_{\rm B}(T/m^{*})\Sigma(m^{*})
\eeq
Here $m^{*}=1$ in the {\it glassy liquid} regime where the complexity is finite $\Sigma(1) > 0$ and one anticipates non-trivial two-step relaxations with $\beta$ and $\alpha$ relaxations: at shorter time scales before the $\alpha$ relaxation takes place the system behaves as a piece of a solid (glass)
with bounded thermal fluctuations parameterized by
$\Delta < \infty$ and/or $q > 0$.
As the density $\hat\varphi$ is increased the complexity $\Sigma(1)$ can vanish
at some $\hat\varphi_{\rm K}$, signaling a thermodynamic glass transition called as Kauzmann transition.
In the ideal glass phase $\hat\varphi > \hat\varphi_{\rm K}$,
$m^{*}$ is defined by vanishing of the complexity $\lim_{m \to m^{*}_{-}}\Sigma(m)=0$
and it decreases with increasing  $\hat\varphi$.
Eventually, at a certain $\hat\varphi_{\rm GCP}$ called as glass close packing density,
$m^{*}$ vanishes signaling jamming (of the ideal glass state) where the pressure diverges.

The analysis should start by examing the lowest density
called as dynamic glass transition density $\hat\varphi_{\rm d}$
beyond which glassy states $\Delta < \infty$ and/or $q>0$ exist.
In the glassy liquid regime  $\hat\varphi_{\rm d} < \hat\varphi < \hat\varphi_{\rm K}$
we have  $m^{*}=1$ and non-trivial solutions can be obtained
by extremizing the Franz-Parisi potential $\nu_{\rm FP}[\hat{\cal Q}]$,
which is defined as,
\beq
-\beta (1+s)\phi_{1+s}[\hat{\cal Q}]=
-\beta \phi_{1}[\hat{\cal Q}]-\beta \nu_{\rm FP}[\hat{\cal Q}]s+O(s^{2})
\label{eq-sm-def-FP}
\eeq

    \subsection{Kauzmann transition and glass close packing}

    At large dimensions $d \gg 1$ we have $c_{\rm nt}$
    given by \eq{eq-sm-cnt} behaves as,
    \beq
    c_{\rm nt}= \frac{d}{2}(1-3m) \ln d
    \eeq
    Here we used $\ln \rho  \simeq  \frac{d}{2}\ln d $ at $d \gg 1$
    which follows from  \eq{eq-sm-vaphi-rho},
    $\Omega_{d}/d=\pi^{d}/\Gamma(1+d/2)$
    and the Stirling's formula $\Gamma(1+z) \simeq z\ln z - z$
    for $z \gg 1$.

Using the above results we have at large densities $\hat\varphi \gg 1$,
\beqn
-\beta m \phi_{m}(\Delta,q)=
d \left( \frac{1}{2}-\frac{3m}{2} \right) \ln d +\frac{d}{2}\hat\varphi
({\cal G}_{m}(0,1)-1)
\eeqn
Then following the standard prescription  \cite{Mo95} we obtain
\beqn
&& -\beta f^{*}(m)=\partial_{m}(-\beta m \phi_{m})=-\frac{3}{2}d \ln d + \frac{d}{2}\hat\varphi \partial_{m} {\cal G}_{m}(0,1) \\
&& \Sigma^{*}(m)=-\beta m \phi_{m}+\beta m f^{*}(m)=\frac{d}{2}\ln d
-\frac{d}{2}\hat\varphi
\left(
1+m^{2}\partial_{m}m^{-1}{\cal G}_{m}(0,1)
\right)
\eeqn

In the glassy liquid regime, the complexity is given by
\beq
\frac{\Sigma(1)}{
  (d/2)\ln d}= 1- \left. \hat\varphi ( 1 + m^{2}\partial_{m} m^{-1}
{\cal G}_{m}(0,1) \right |_{m=1}
\eeq
which implies the Kauzmann transition takes place as
\beq
\frac{\hat\varphi_{\rm K}}{\ln d}=
\frac{1}{
  \left.
 1 + m^{2}\partial_{m} m^{-1}
{\cal G}_{m}(0,1) \right |_{m=1}
}
\label{eq-sm-phi-K}
\eeq

In the ideal glass state at $\hat\varphi > \hat\varphi_{\rm K}$,
we choose $m=m^{*}$ such that $\Sigma(m^{*})=0$. The glass close packing
density at which the ideal glass state exhibit jamming
is obtained by $\Sigma(m^{*})=0$. This implies the glass close packing density,
\beq
\frac{\hat\varphi_{\rm GCP}}{\ln d}=1
\label{eq-sm-phi-GGP}
\eeq

\section{Patchy colloid}

\subsection{Replicated free-energy: 1RSB ansatz}

For patchy colloid we consider the Kern-Frenkel potential   \cite{kern2003fluid}
which is given by,
\beq
e^{-\beta V(\xi,x,x')} =\theta(\xi)
 +(1-e^{1/\hat{T}}) [\theta(\xi-\hat{\sigma})-\theta(\xi)]\Omega(x,x'). \qquad
\eeq
where 
\beq
\Omega(x,x') =(\theta(-x-\delta_{+})+\theta(x-\delta_{-}))(\theta(x'-\delta_{+})+\theta(-x'-\delta_{-}))
\label{eq-sm-omega-2-patch}
\eeq
for the system of two patches at head and tail (See Fig.~1).

Thus compared with the generic  potential $V(\xi,x,x',h)$, the Kern-Frenkel potential
does not depend on $h$, which represents the inner product of the directors $\vS_{1}\cdot\vS_{2}$.

Fortunately we find the function $g(\xi,x,x')$ in \eq{eq-sm-g-integral-form} can be obtained analytically as,
\beq
g(\xi,x,x') =
\Theta\left( \frac{\xi}{\sqrt{2\Delta}} \right)
 +(1-e^{1/\hat{T}})
 \left[
   \Theta\left( \frac{\xi-\hat\sigma}{\sqrt{2\Delta}} \right)- \Theta\left( \frac{\xi}{\sqrt{2\Delta}} \right)
\right]
\hat{\Omega}(x,x')
\label{eq-sm-g-patch-colloid}
\eeq
\beq
\nonumber
\eeq
with
\beqn
\hat{\Omega}(x,x')&=&\int Dz_{x}Dz_{x'} \Omega(x-\sqrt{1-q}z_{x},x'-\sqrt{1-q}z_{x'})\nonumber \\
&=& 
\left[\Theta \left(\frac{-x-\delta_{+}}{\sqrt{2(1-q)}} \right)+\Theta \left( \frac{x-\delta_{-}}{\sqrt{2(1-q)}}\right) \right]
\left[\Theta \left( \frac{x'-\delta_{+}}{\sqrt{2(1-q)}} \right)+\Theta \left( \frac{-x'-\delta_{-}}{\sqrt{2(1-q)}} \right)\right]
\label{eq-sm-omega-2-patch2}
\eeqn
In the above equations we used a function $\Theta(x)$,
\beq
\Theta(x) \equiv \int_{-\infty}^{x}\frac{dz}{\sqrt{\pi}}e^{-z^{2}}
=\gamma_{1/2} \otimes \theta(x)=\frac{1}{2} (1+{\rm erf}(x)),
\label{eq-sm-def-Theta}
\eeq
with ${\rm erf}(x)$ being the error function,
\beq
{\rm erf}(x)=\frac{2}{\sqrt{\pi}}\int_{0}^{x} dy e^{-y^{2}}=-{\rm erf}(-x),
\eeq

In order to evaluate the free-energy \eq{eq-sm-replica-free-energy-1RSB} and \eq{eq-sm-FP-1RSB}
we are left with integrals over $\xi$, $z_{x}$ and $z_{x'}$ which we perform numerically.

%\subsection{Emergence of glassy states}
\subsection{Kauzmann transition and glass close packing}

The function ${\cal G}_{m}(\Delta,q)$ defined in \eq{eq-sm-def-cal-G} is obtained as,
\beqn
&& {\cal G}_{m}(\Delta,q)= \int_{-\infty}^{\infty}d\xi e^{\xi}
\left. \left(
\int {\cal D}z_{x}{\cal D}z_{x'}
g^{m}(\Xi) \right |_{\Xi=(\xi+\Delta/2,\sqrt{q}z_{x},\sqrt{q}z_{x'})} -\theta(\zeta) \right)  \nonumber \\
&& = \int_{-\infty}^{\infty}d\xi e^{\xi}
%\left.
\Biggl \{
%\left.
\int {\cal D}z_{x}{\cal D}z_{x'}
\Biggl[
  \Theta\left( \frac{\xi+\Delta/2}{\sqrt{2\Delta}} \right) 
+(1-e^{1/\hat{T}}) \Bigl[ \Theta\left( \frac{\xi+\Delta/2-\hat\sigma}{\sqrt{2\Delta}} \right)
- \Theta\left( \frac{\xi+\Delta/2}{\sqrt{2\Delta}} \right)
\Bigr]\hat{\Omega}(\sqrt{q}z_{x},\sqrt{q}z_{x'})
%\Theta(t)
%+(1-e^{1/\hat{T}}) \left[ \Theta(t-\frac{\hat\sigma}{\sqrt{2\Delta}})
%- \Theta(t)
%\right]\hat{\Omega}(x,x')
\Biggr]^{m}
%\right |_{\xi=\xi+\Delta/2}
%\right.
\nonumber \\
&&
%\left. 
%g^{m}(\Xi)
\hspace*{10cm}
-\theta(\zeta) \Biggr \}
%\right |_{\Xi=(\xi+\Delta/2,\sqrt{q}z_{x},\sqrt{q}z_{x'})}
\eeqn
where $\hat\Omega(x,x')$ is defined in \eq{eq-sm-omega-2-patch2}.
This yields,
\beqn
   {\cal G}_{m}(0,1)&=&\int_{-\infty}^{\infty} d\xi e^{\xi}
    \left\{
    \int {\cal D}z_{x}{\cal D}z_{x'}
    \left[
        [1-(1-e^{1/\hat{T}})\Omega(z_{x},z_{x'})]\theta(\zeta)
        +(1-e^{1/\hat{T}})\Omega(z_{x},z_{x'})\theta(\zeta-\sigma)
        \right]^{m}-\theta(\zeta)
    \right\} \nonumber \\
    & =&   (e^{\hat\sigma}-1)(e^{m/\hat{T}}-1)\overline{\Omega}
    \eeqn
    where we introduced
    \beq
        \overline{\Omega}=\int {\cal D} z_{x}{\cal D} z_{x'}
    \hat{\Omega}(z_{x},z_{x'})  
    \eeq
In particular for the two patch system \eq{eq-sm-omega-2-patch2} we find,
\beq
\overline{\Omega}= (\Theta(-\delta_{+}/\sqrt{2})+\Theta(-\delta_{-}/\sqrt{2}))^{2}
\eeq

From the above results we find the Kauzmann transition density
\eq{eq-sm-phi-K}, 
\beq
\frac{\hat\varphi_{\rm K}}{\ln d}=
\frac{1}{
  1 + 
(e^{\hat{\sigma}}-1)
[1+(1/\hat{T}-1)e^{1/\hat{T}}]\overline{\Omega}
}
\eeq
while the glass close packing density $\hat\varphi_{\rm GCP}$
is given by \eq{eq-sm-phi-GGP}.

%\subsection{Janus particles}

%\subsection{``Obi'' particles}

%\subsection{Head tail symmetry and continuous transition}
%
%    In the case of head-tail symmetric 2 patchy colloid, the onset of the
%    orientational glass order parameter is continuous and decouples from the
%    translational glass transition.
%Here we show we how the asymmetry changes the situation.
%
%\begin{figure}[h]
% \includegraphics[width=0.8\textwidth]{data/2patch/Assymetric/fig_head_tail_anisotropy.pdf}
% \caption{
%   The behaviour of the glass order parameters with/without head-tail symmetry.
%   The head-tail ratio decreases as $\delta_{+}/\delta_{-}=1.0,0.9,0.8,0.7,0.6,0.5$  in the direction indicated by the arrows.
%   Here $\hat{T}=0.5$,$\sigma=0.1$ and $\delta_{-}=0.8$.
% }
%  \label{fig_head_tail_anisotropy}
%\end{figure}
%
%\subsection{One patch cases}
%
%For the case of one-patch (one patch just at the head)
%we just need to consider $\delta_{+}=\delta$ and $\delta_{-}=\sqrt{d} \to \infty$
%which yields $\Omega(x,x')=\theta(-x-\delta)\theta(x'-\delta)$.
%
%Note that usual sticky colloid is recovered by
%$\delta= - \sqrt{d}\to -\infty$.
%

%\section{Reversed 2-patchy colloid}

%\begin{figure}[h]
%% \includegraphics[width=0.7\textwidth]{fig_.pdf}
%  \caption{
% }
%  \label{fig_}
%\end{figure}

%\bibliographystyle{mioaps}
%\bibliography{HS,ref_yoshino}

%\end{document}

\end{document}